\documentclass[aps,twocolumn,superscriptaddress,floatfix,nofootinbib]{revtex4-1}
\usepackage{graphicx,amsmath,amssymb,verbatim,color}
\usepackage{booktabs}
\usepackage[utf8]{inputenc}
\usepackage{fancyhdr, fancybox, empheq} 
\usepackage{color}

\usepackage[colorlinks=true,citecolor=blue,linkcolor=blue,urlcolor=blue, backref=false,pdfborder={0 0 0}]{hyperref}

\newcommand{\beq}{\begin{equation}}
\newcommand{\eeq}{\end{equation}}
\newcommand{\barr}{\begin{eqnarray}}
\newcommand{\earr}{\end{eqnarray}}

\newcommand{\aapr}{A\&ARv}
\newcommand{\sovast}{Soviet Astronomy}
\newcommand{\mnras}{MNRAS}
\newcommand{\apjl}{ApJL}

\newcommand{\rme}{\textrm{e}}
\newcommand{\bs}{\boldsymbol}
\newcommand{\ho}{\hat{\Omega}}
\newcommand{\hp}{\hat{p}}
\newcommand{\hq}{\hat{q}}
\newcommand{\bsm}[1]{\bs{\mathcal{#1}}}

\begin{document}

\title{Fisher formalism for anisotropic gravitational-wave background searches with pulsar timing arrays}

\author{Yacine Ali-Ha\"imoud}
\affiliation{Center for Cosmology and Particle Physics, Department of Physics,
New York University, New York, NY 10003, USA}

\author{Tristan L.~Smith}
\affiliation{Department of Physics and Astronomy, Swarthmore College, 500 College Ave., Swarthmore, PA 19081, USA}

\author{Chiara M.~F.~Mingarelli}
\affiliation{Center for Computational Astrophysics, Flatiron Institute, 162 Fifth Ave, New York, NY 10010, USA}
\affiliation{Department of Physics, University of Connecticut, 196 Auditorium Road, U-3046, Storrs, CT 06269-3046, USA}
\date{\today}

\begin{abstract}

Pulsar timing arrays (PTAs) are currently the only experiments directly sensitive to gravitational waves with decade-long periods. Within the next five to ten years, PTAs are expected to detect the stochastic gravitational-wave background (SGWB) collectively sourced by inspiralling supermassive black hole binaries. It is expected that this background is mostly isotropic, and current searches focus on the monopole part of the SGWB. Looking ahead, anisotropies in the SGWB may provide a trove of additional information both on known and unknown astrophysical and cosmological sources. In this paper, we build a simple yet realistic Fisher formalism for anisotropic SGWB searches with PTAs. Our formalism is able to accommodate realistic properties of PTAs, and allows simple and accurate forecasts. We illustrate our approach with an idealized PTA consisting of identical, isotropically distributed pulsars. In a companion paper, we apply our formalism to current PTAs and show that it can be a powerful tool to guide and optimize real data analysis.

\end{abstract}

\maketitle

\section{Introduction}

The promise of timing pulsars to detect nHz gravitational waves (GWs) was pointed out more than four decades ago \cite{Sazhin_78, Detweiler_79}, and the application to a \emph{stochastic} gravitational-wave background (SGWB) was studied shortly after \cite{Hellings:1983fr}. In that seminal paper, Hellings and Downs derived the response of pulsar timing residual correlations to an isotropic SGWB, and were the first to combine several pulsars to extract upper limit on the SGWB amplitude. Since then, several collaborations \cite{EPTA, Nanograv, PPTA, IPTA} have been timing arrays of pulsars, some for over two decades, and getting increasingly stringent upper limit on the SGWB amplitude \cite{Lentati_15, Shannon_15, Arzoumanian_18}. If our understanding of galaxy formation and merger history is correct, pulsar timing arrays (PTAs) should detect the SGWB generated by inspiralling supermassive black hole binaries (SMBHBs) within the next decade \cite{Taylor+:2016, Burke-Spolaor:2018bvk, Mingarelli2019}. In addition to this astrophysical background, other, more exotic processes could also contribute to the nHz SGWB, see e.g.~Refs.~\cite{siemensetal2019, Burke-Spolaor:2018bvk, lms+16}.

While most current searches assume perfect isotropy, there is likely some level of anisotropy in the SGWB\footnote{Given that the SGWB is the power spectrum of the gravitational-wave strain, we are technically referring to \emph{statistical} anisotropy. However, since the gravitational-wave strain itself is \emph{necessarily} anisotropic (the only isotropic rank-2 and trace-free tensor is the null tensor), we will drop the qualifier ``statistical" when referring to anisotropies of the (scalar) SGWB intensity, as there is no risk of confusion.}. For one, the finite number of SMBHBs should inevitably imply some level of anisotropy \cite{Cornish_13, Mingarelli:2013dsa, Mingarelli_17}. On large scales, one also expects the distribution of SMBHBs to trace cosmic structure \cite{Burke-Spolaor:2018bvk}. Independently of their physical origin, it is important to understand what kind and what level of SGWB anisotropy PTAs might in principle be able to detect. Indeed, such anisotropy might eventually prove a powerful discriminant between different models of the SGWB.

The standard approach to study the detectability of SGWB anisotropies has been to harness the full power of Bayesian analysis pipelines used for real data \cite{Taylor:2013esa, Taylor:2015udp}. While this approach provides the most accurate results, it is computationally demanding, and does not allow for making quick estimates nor building intuition. Recently, Ref.~\cite{Hotinli_19} developed a simplified approach, holding for an idealized PTA constituted of a large number of identical, isotropically-distributed pulsars\footnote{This underlying assumption is not explicitly spelled out in Ref.~\cite{Hotinli_19}, but is required for the harmonic transform of the timing residuals to be uncorrelated and have $\ell$-independent noise, as assumed there.}. In this paper, we fill the gap between these two methods, by deriving a simple, yet realistic, Fisher formalism for SGWB anisotropies. We moreover break away from the spherical-harmonic decomposition of the SGWB that most past works have relied on thus far~\citep{Mingarelli:2013dsa, Gair:2014rwa, Taylor:2015udp, Taylor_20}, as we argue it is poorly adapted to real PTAs (see Ref.\cite{Cornish:2014rva} for a different mapping approach, in the case of continuous waves). Our formalism allows for real pulsar distributions and noise properties, and yet permits us to make quick detectability estimates without running time-consuming Monte Carlo Markov Chains. While our formalism is not a substitute for a full-on data analysis, it provides useful tools to make \emph{forecasts}, as well as \emph{guide} and \emph{optimize} SGWB searches with PTAs. Our philosophy is inspired by what has long been the norm in the field of cosmology, where Fisher analyses (e.g.~\cite{Knox_95, Tegmark_00}) are routinely used and have proven extremely useful, not only to produce forecasts, but also to make the field accessible to a broader community.

This work is the first of a series of two articles. In the present paper, we expound the theoretical formalism, culminating in the derivation of the PTA Fisher matrix for an anisotropic SGWB, Eq.~\eqref{eq:Fisher-final}. Along the way, we rederive some known results with a fresh approach, using only frame-independent, geometric expressions. In the companion paper (hereafter, Paper II), we shall apply this tool to several practical examples, and illustrate how to make forecasts or optimize searches for SGWB anisotropies.

This paper is organized as follows. In Section \ref{sec:response}, we start by describing the statistical properties of the SGWB in a new frame-independent, geometric fashion, and then derive the response of a timing array to the SGWB. In Section \ref{sec:Fisher}, we derive the Fisher matrix for the SGWB intensity, which is our main result. In Section \ref{sec:dense-Fisher}, we apply our results to the idealized case of a dense array of identical pulsars isotropically distributed on the sky. We presents a new calculation of the Hellings and Downs curve in Appendix \ref{app:HD}, derive the covariance matrix of time residual bandpowers in Appendix \ref{app:radiometer}, and the Fisher matrix in the limit of a large number of identical, isotropically distributed pulsars in Appendix \ref{app:dense-limit}. Throughout we use units in which the speed of light is unity. A summary of our notation can be found in Table \ref{tab:notation}.

\renewcommand{\arraystretch}{1.4}
\begin{center}
\begin{table*}
\begin{tabular}{| c | c | c | c |}
\hline 
Symbol & Description & Dimensions & Defining equation\\
\hline \hline
$\bs{1}(\ho)$ & isotropic map equal to unity for all directions $\ho$ & dimensionless & $\bs{1}(\ho) = 1 \ \ \ \forall ~\ho$\\
$\mathcal{C}_{IJ}$ & covariance of estimators of time-residual band-power & time$^4$ & $\mathcal{C}_{IJ} \equiv \textrm{cov}(\widehat{\mathcal{R}}_I, \widehat{\mathcal{R}}_J)$ \\
$\delta_{ab}^{\bot \ho}$ & identity tensor in the plane orthogonal to $\ho$ & dimensionless & \eqref{eq:delta_perp}\\ 
$\Delta t_p$ & observation cadence of pulsar $p$ & time & \\
$\Delta f$ & frequency bandwidth for bandpowers & frequency & \\
$f$ & gravitational-wave frequency & frequency & \eqref{eq:Fourier}\\
$\bsm{F}(\ho, \ho')$ & Fisher matrix of band-integrated GW intensity & dimensionless & \eqref{eq:Fisher-def}\\
$F(\ho, \ho')$ & reduced Fisher matrix for identical pulsars & dimensionless & \eqref{eq:reduced-F}\\
$F_{IJ}$ & $N_{\rm pair} \times N_{\rm pair}$ discretized reduced Fisher matrix & dimensionless & \eqref{eq:FIJ}\\
$\bs{\gamma}_{\hp \hq}(\ho) = \bs{\gamma}_I(\ho)$ & pairwise timing response function at pulsar pair $I = (p, q)$ & dimensionless &\eqref{eq:Gamma}\\
$\bs{\gamma}_I^*(\ho)$ & dual map of $\bs{\gamma}_I(\ho)$ & dimensionless &$\bs{\gamma}_I^* \cdot \bs{\gamma}_J = \delta_{IJ}$ \\ 
$h_{ab}(t, \ho)$ & gravitational-wave strain & dimensionless & \eqref{eq:metric}\\
$h_{ab}(f, \ho)$ & Fourier transform of $h_{ab}(t)$ & 1/frequency & \eqref{eq:Fourier}\\
$h_c(f)$ & characteristic gravitational-wave strain & dimensionless & \eqref{eq:hc-I} \\
$\mathcal{H}(\mu)$ & Hellings and Downs function (response to an isotropic SGWB)& dimensionless &\eqref{eq:HD} \\
$I, J$ & labels of unique pulsar pairs & indices &  $I = (p, q)$, $J = (p', q')$ \\
$\bsm{I}(f, \ho)$ & total intensity of the SGWB & 1/frequency & \eqref{eq:Pabab-to-I}\\
 $\bsm{I}_f(\ho)$ or $\bsm{I}(\ho)$ & band-integrated SGWB intensity & dimensionless & \eqref{eq:I_f}\\
  $\mathfrak{I}_{abcd}(\ho)$ & geometric dependence of the SGWB total intensity piece &dimensionless &\eqref{eq:Iijkl} \\
  $\mathcal{L}_{abcd}(f, \ho)$ & linear polarization tensor of the SGWB & 1/frequency & \eqref{eq:Pijkl2}\\
  $N_p(t)$ & intrinsic time-residual noise of pulsar $p$ & time & \eqref{eq:total-Rp}\\
$N_p(f)$ & Fourier transform of $N_p(t)$ & time/frequency & \\
 $\ho$ &  gravitational-wave direction of propagation & unit vector & \eqref{eq:Fourier}\\
 $p, q$ & labels of individual pulsars & indices & \\
 $\hp, \hq$ & unit vectors pointing in the direction of individual pulsars & unit vectors & \\
 $\bsm{P}_{abcd}(f, \ho)$ & rank-4 power spectrum of the SGWB & 1/frequency & \eqref{eq:Pijkl} \\ 
 $R_{p}^{\rm GW}(t)$ & gravitational-wave-induced time residual of pulsar p  &  time & \eqref{eq:R}\\
 $R_{p}^{\rm GW}(f)$ & Fourier transform of $R_p^{\rm GW}(t)$ &  time/frequency & \eqref{eq:R-Fourier}\\
 $R_{p}(t)$ & total time residual of pulsar p  &  time & \eqref{eq:total-Rp}\\
 $R_{p}(f)$ & Fourier transform of $R_p(t)$ &  time/frequency & \\
 $\mathcal{R}_I(f) = \mathcal{R}_{pq}(f)$ & cross-power spectrum of time residuals of pulsar pair $I = (p, q)$ & time$^2$/frequency &\eqref{eq:R_pq-def}\\
 $\mathcal{R}_{I, f}$ or $\mathcal{R}_I$ & band-integrated time-residual power spectrum & time$^2$ & \eqref{eq:I-to-R} \\
 $\widehat{\mathcal{R}}_I$ & estimator for $\mathcal{R}_I$ & time$^2$ & \\
 $\sigma_p^2(f)$ & pulsar timing noise power spectrum & time$^2$/frequency& \eqref{eq:sigma_p^2}\\
 $\sigma_{p, f}^2$ or $\sigma_p^2$ & band-integrated pulsar timing noise & time$^2$ & \eqref{eq:sigma_p-band}\\
 $\mathcal{T}_p(f)$ & timing-model-fitting transmission function of pulsar $p$ & dimensionless &\eqref{eq:Fisher-final}\\ 
 $T_p$ & total observation time of pulsar $p$ & time & \\
 $T_{pq}$ & effective total observation time of pulsar pair $p, q$& time & $T_{pq} = \min(T_p, T_q)$\\
 $\bsm{V}(f, \ho)$ & circular polarization amplitude of the SGWB & 1/frequency & \eqref{eq:V} \\  
 $\mathfrak{V}_{abcd}(\ho)$ & geometric dependence of the SGWB circular polarization piece & dimensionless &\eqref{eq:Vijkl} \\
 $\bsm{Y}_{\ell m}(\ho)$ & real spherical harmonics & dimensionless & \\
\hline
\end{tabular}
\caption{Summary of the notation used in this paper, in alphabetic order, with the defining equations.} \label{tab:notation}
\end{table*}
\end{center}

\section{Geometric description of the response of pulsar pairs to a SGWB} \label{sec:response}

\subsection{Geometric decomposition of the SGWB power spectrum}

In this section we present a new, geometric and frame-independent decomposition of the SGWB power spectrum. In Sec \ref{sec:frame-dep}, we relate the new expressions, Eqs. \eqref{eq:Pijkl2}, \eqref{eq:Iijkl} and \eqref{eq:Vijkl}, to those commonly found in the PTA literature. Our frame-independent expressions will prove very powerful later on as they allow us to express all relevant observables through explicit functions of scalar products between unit vectors.  

We decompose the GW strain $h_{ab}(t, \vec{x})$ in the Fourier domain as follows 
\beq
h_{ab}(t, \vec{x}) = \int_{-\infty}^\infty df \int d^2 \ho~  h_{ab}(f, \ho)~ e^{2i \pi f (t - \ho \cdot \vec{x})}, \label{eq:Fourier}
\eeq
where ${h}_{ab}(f, \ho)$ is symmetric, trace-free, and transverse to the direction of propagation $\ho$, i.e.~$\ho^a h_{ab}(f, \ho) = 0$.

If we assume that the SGWB is a stationary Gaussian random field (as would be the case if it is generated by a large number of uncorrelated sources), it is entirely determined by its power spectrum $\bsm{P}_{abcd}$, which we normalize as follows:
\barr
\langle h_{ab}(f, \ho) h_{cd}^*(f', \ho') \rangle &=& \frac{\delta_{\rm D}(\ho', \ho)}{4 \pi} \frac{\delta_{\rm D}(f'-f)}{2}\nonumber\\
&&\times \bsm{P}_{abcd}(f, \ho), \label{eq:Pijkl}
\earr
where $\delta_{\rm D}$ is the Dirac function. The Dirac function in frequency stems from time-translation invariance (i.e., stationarity) of the correlation function $\langle h_{ab}(t) h_{cd}(t + \Delta t)\rangle$, and the angular Dirac function $\delta_{\rm D}(\ho', \ho)$ stems from spatial-translation invariance (i.e., statistical isotropy). 

The definition \eqref{eq:Pijkl} implies the following hermiticity property:
\beq
\bsm{P}_{cdab}(f, \ho) = \bsm{P}_{abcd}^*(f, \ho). \label{eq:hermiticity}
\eeq
In addition, the reality of $h_{ab}(t, \vec{x})$ implies that $h_{ab}(-f, \ho) = h_{ab}^*(f, \ho)$, in turn implying
\beq
\bsm{P}_{abcd}(- f, \ho) = \bsm{P}_{abcd}^*(f, \ho). \label{eq:reality}
\eeq
The GW power spectrum $\bsm{P}_{abcd}$ is a rank-4 tensor, which is symmetric and trace-free for the first and last pair of indices, and transverse to $\ho$ in each index. Hence it has 4 independent components, which are {\it a priori} complex. 
The hermiticity property \eqref{eq:hermiticity} reduces the number of independent component to 4 \emph{real} quantities. This is the same number of components as the (rank-2) electromagnetic intensity tensor, i.e.~the power spectrum of the electromagnetic field. Just like the latter, we may decompose $\bsm{P}_{abcd}$ into a component proportional to its trace (the total intensity)
\beq
\bsm{I}(f, \ho) \equiv \frac14 \bsm{P}_{abab}(f, \ho), \label{eq:Pabab-to-I}
\eeq
a component proportional to a real pseudo-scalar $\bsm{V}(f, \ho)$ (the circular polarization), and a real, fully trace-free linear-polarization tensor $\bsm{L}_{abcd}(f, \ho)$, carrying the two remaining independent components. More specifically, we want to decompose the power spectrum as follows:
\barr
\bsm{P}_{abcd}(f, \ho) &=& \bsm{I}(f, \ho) ~\mathfrak{I}_{abcd}(\ho) + i \bsm{V}(f, \ho) ~\mathfrak{V}_{abcd}(\ho)\nonumber\\
&& + \bsm{L}_{abcd}(f, \ho), \label{eq:Pijkl2}
\earr
where the (real) tensors $\mathfrak{I}_{abcd}(\ho)$ and $\mathfrak{V}_{abcd}(\ho)$ are purely geometric, frequency-independent objects. The traces of the tensors appearing in Eq.~\eqref{eq:Pijkl2} are
\beq
\mathfrak{I}_{abab} = 4, \ \ \ \ \ \mathfrak{V}_{abab} = 0, \ \ \ \ \bsm{L}_{abad} = 0.
\eeq
The reality condition \eqref{eq:reality} implies that $\bsm{I}(f, \ho)$ and $\bsm{L}_{abcd}(f, \ho)$ are even functions of $f$, while $\bsm{V}(f, \ho)$ is an odd function of $f$.

The geometric objects $\mathfrak{I}_{abcd}(\ho)$ and $\mathfrak{V}_{abcd}(\ho)$ must be built exclusively out of isotropic tensors, i.e.~the Kroneker delta and the Levi-Civita tensor $\epsilon_{abc}$, and of $\ho$, which is the only preferred direction. For $\bsm{I}$ to be a real scalar, the tensor $\mathfrak{I}_{abcd}$ must only contain Kronecker deltas and $\ho$, i.e.~be of the form 
\beq
\mathfrak{I}_{abcd}(\ho) \propto \delta_{ab} \delta_{cd}, ..., \delta_{ab} \ho_c \ho_d,..., \ho_a \ho_b \ho_c \ho_d,
\eeq
where the ... include all permutations of indices. By imposing that $\mathfrak{I}_{abcd}$ has the symmetries of $\bsm{P}_{abcd}$, and, from Eq.~\eqref{eq:hermiticity}, is symmetric under exchange of the first and last pair of indices, one finds that the \emph{only} rank-4 tensor satisfying these properties, with the appropriate normalization $\mathfrak{I}_{abab} = 4$ is
\beq
\mathfrak{I}_{abcd}(\ho) = \delta^{\bot \ho}_{ac} \delta^{\bot \ho}_{bd} + \delta^{\bot \ho}_{ad} \delta^{\bot \ho}_{bc} - \delta^{\bot \ho}_{ab} \delta^{\bot \ho}_{cd},   \label{eq:Iijkl}
\eeq
where $\delta^{\bot \ho}_{ab}$ is the identity tensor projected on the plane orthogonal to $\ho$,
\beq
\delta^{\bot \ho}_{ab} \equiv \delta_{ab} - \ho_a \ho_b. \label{eq:delta_perp}
\eeq
For $\bsm{V}$ to be a pseudo-scalar, the geometric object $\mathfrak{V}_{abcd}(\ho)$ must be proportional to $\ho^a \epsilon_{abc}$, and otherwise be built out of Kronecker deltas and $\ho$. It must have the same symmetry properties as $\bsm{P}_{abcd}$ and satisfy $\mathfrak{V}_{cdab} = - \mathfrak{V}_{abcd}$. Up to a proportionality constant (which we chose in order to match existing derivations, as we will see shortly), the only possible tensor with the appropriate symmetry properties is 
\beq
\mathfrak{V}_{abcd}(\ho) = \ho^e \left(\epsilon_{ea(c} \delta^{\bot \ho}_{d)b} + \epsilon_{eb(c} \delta^{\bot \ho}_{d)a} \right), \label{eq:Vijkl}
\eeq
where $X_{(ab)} \equiv (X_{ab} + X_{ba})/2$ represents symmetrization. With this convention, we have $\mathfrak{V}_{abad}  = 2 \ho^e \epsilon_{ebd}$, thus the amplitude of circular polarization $\bsm{V}$ can be obtained from 
\beq
\bsm{V} = \frac1{4i} \bsm{P}_{abad} ~\epsilon_{bde}~\ho^e. \label{eq:V}
\eeq
Finally, $\bsm{L}_{abad}$ contains information about the \emph{linear} polarization of the SGWB. 

To conclude, Eqs. \eqref{eq:Pijkl2}, \eqref{eq:Iijkl} and \eqref{eq:Vijkl} form a geometric, frame-independent decomposition of the GW power spectrum, with the most general frequency and angular dependence. In the remainder of this paper, we will focus on the total-intensity part of the SGWB, i.e.~assume that the circular and linear polarization components are subdominant (it is conceptually straightforward to generalize our formalism to a polarized SGWB). In the majority of works on the SGWB, the intensity is assumed to be isotropic, $\bsm{I}(f, \ho) = \mathcal{I}(f)$. In this case, the SGWB intensity is just half of the one-sided GW strain spectral density $S_h(f) = h_c^2(f)/f$, where $h_c(f)$ is the characteristic strain \cite{Phinney:2001di}. More generally, these quantities are related to the angle-average of $\bsm{I}(f, \ho)$ through
\beq
S_h(f) = \frac{h_c^2(f)}{f} = 2\int \frac{d \ho}{4 \pi} ~\bsm{I}(f, \ho), \label{eq:hc-I}
\eeq
as can be seen from taking the trace of Eq.~\eqref{eq:Pijkl} and integrating it over angles, and comparing to, e.g., Ref.~\cite{Jaffe_02}. The authors of Ref.~\cite{Mingarelli:2013dsa} consider the possibility of an anisotropic GW intensity, with a factorized frequency and angular dependence. Their convention corresponds to $\bsm{I}(f, \ho) = 8 \pi H(f) P(\ho)$. A similar convention is adopted (up to a factor of 2) in Ref.~\cite{Gair:2014rwa}.

\subsection{Connection with standard frame-dependent notation} \label{sec:frame-dep}

We now relate the geometric, frame-independent description given above to the more standard frame-dependent expressions found in the literature. 
For a given direction of GW propagation $\ho$, one may pick two arbitrary vectors $\hat{m}$ and $\hat{n} = \ho \times \hat{m}$ orthogonal to $\ho$, and define the two polarization basis tensors 
\beq
e_{ab}^+ \equiv \hat{m}_a \hat{m}_b - \hat{n}_a \hat{n}_b, \ \ \ \ e_{ab}^\times \equiv \hat{m}_a \hat{n}_b + \hat{n}_a \hat{m}_b.
\eeq
Since the triad $\hat{m}, \hat{n}, \ho$ forms an orthonormal basis, we have 
\barr
\hat{m}_a \hat{m}_b + \hat{n}_a \hat{n}_b &=& \delta_{ab} - \ho_a \ho_b = \delta_{ab}^{\bot \ho},\\
\hat{m}_b \hat{n}_d - \hat{n}_b \hat{m}_d &=& \ho^e \epsilon_{ebd},
\earr
independent of the choice of $\hat{m}, \hat{n}$. From these expressions, one can show that the tensors $\mathfrak{I}_{abcd}$ and $\mathfrak{V}_{abcd}$ defined in Eqs.~\eqref{eq:Iijkl} and \eqref{eq:Vijkl} are given by 
\barr
\mathfrak{I}_{abcd} &=& e_{ab}^+ e_{cd}^+ + e_{ab}^\times e_{cd}^\times , \label{eq:Iabcd-2}\\
\mathfrak{V}_{abcd} &=& e_{ab}^+ e_{cd}^\times - e_{ab}^\times e_{cd}^+. \label{eq:Vabcd-2}
\earr
Let us now project the strain onto the basis $e_{ab}^+, e_{ab}^\times$:
\beq
h_{ab}(f, \ho) = h_+(f, \ho) e_{ab}^+ + h_\times(f, \ho) e_{ab}^\times.
\eeq
For the sake of compactness, for any two stochastic variables $X, Y$, we define the quantity $\langle X Y^* \rangle'$ such that
\barr
\langle X(f, \ho) Y^*(f', \ho') \rangle = \frac{\delta_{\rm D}(\ho', \ho)}{4 \pi} \frac{\delta_{\rm D}(f'-f)}{2} ~ \langle X Y^* \rangle'.~~~~~
\earr
In words, $\langle X Y^* \rangle'$ is the cross-power spectrum of $X$ and $Y$. With this convention, the GW power spectrum is 
\barr
\bsm{P}_{abcd} &=& \langle h_{ab} h_{cd}^* \rangle' \nonumber\\
&=&  \langle |h_+|^2 \rangle' e_{ab}^+ e_{cd}^+ +  \langle h_+ h_\times^* \rangle' e_{ab}^+ e_{cd}^\times  \nonumber\\
&+&  \langle h_\times h_+^* \rangle' e_{ab}^\times e_{cd}^+ + \langle |h_\times|^2 \rangle' e_{ab}^\times e_{cd}^\times. 
\earr
Now using Eqs.~\eqref{eq:Iabcd-2} and \eqref{eq:Vabcd-2}, we see that we can write the GW power spectrum in the form \eqref{eq:Pijkl2}, with
\barr
\bsm{I} &=& \frac12 \langle |h_+|^2 + |h_\times|^2 \rangle',\\
\bsm{V} &=&  \frac1{2i} \langle h_+ h_\times^* - h_\times h_+^* \rangle' = \textrm{Im} \langle h_+ h_\times^*\rangle',\\
\bsm{L}_{abcd} &=&  \bs{Q}(e_{ab}^+ e_{cd}^+ - e_{ab}^\times e_{cd}^\times)  +   \bs{U} (e_{ab}^+ e_{cd}^\times + e_{ab}^\times e_{cd}^+), \\  
\bs{Q} &\equiv&  \frac12 \langle |h_+|^2 - |h_\times|^2\rangle', \\
\bs{U} &\equiv& \frac1{2} \langle h_+ h_\times^* + h_\times h_+^* \rangle' =  \textrm{Re}\langle h_+ h_\times^* \rangle' \, .
\earr
These relations clearly show the analogy with the standard Stokes parameters of electromagnetic waves (see e.g.~\cite{Rybicki}). Our normalization matches precisely that of Ref.~\cite{Smith_17} -- see Ref.~\cite{Kato_16} for similar expressions, with a different normalization. 

\subsection{Concise derivation of the timing residuals from GWs}

A common derivation of the time-residual induced by GWs consists of deriving expressions for a null geodesic in the presence of gravitational plane-waves using Killing vectors \cite{Burke_75, 1975GReGr...6..439E, Detweiler_79}. Here we provide a new and concise derivation in the spirit of the first calculation by Ref.~\cite{Sazhin_78}. Our derivation has the advantage of not being limited to a plane wave, but directly applies to a generic superposition of waves, with no special symmetries hence no Killing vector fields. Consider null geodesics in the metric 
\beq
ds^2 = - dt^2 + (\delta_{ab} + h_{ab}) dx^a dx^b, \label{eq:metric}
\eeq
where the GW strain $h_{ab}(t, \vec{x})$ is symmetric, trace-free and transverse ($\partial^a h_{ab} = 0)$.
Specifically, consider light rays originating at a pulsar $p$ (event $P$) and received on Earth (event $E$). We define $d\ell^2 = \delta_{ab} dx^a dx^b$. The null geodesic condition implies that
\barr
dt &=& \left(d \ell^2 + h_{ab} dx^a dx^b\right)^{1/2} = d\ell \left(1 +  h_{ab} \frac{dx^a}{d\ell} \frac{dx^b}{d \ell}\right)^{1/2} \nonumber\\
&=& d \ell \left(1 + \frac12 h_{ab} \frac{dx^a}{d\ell} \frac{dx^b}{d \ell}\right) + \mathcal{O}(h^2), \label{eq:dt-dl}
\earr
where we have expanded to linear order in $h_{ab}$. At this order, we only need to compute $dx^a/d\ell$ along unperturbed geodesics. For unperturbed geodesics traveling along the direction $-\hp$ (so that the unit vector $\hp$ points from Earth to the pulsar), we have $dx^a/d \ell = - \hp^a$. Integrating Eq.~\eqref{eq:dt-dl}, we therefore get
\beq
t_E - t_P = \ell_E - \ell_P + \frac12 \hp^a \hp^b \int_{t_P}^{t_E} d t ~h_{ab}(t, \vec{x}(t)), \label{eq:t-res1}
\eeq
where we have substituted $d \ell$ by $dt$ in the integral, as they are equal to zero-th order in $h_{ab}$, and $\vec{x}(t)$ is the spatial position along the geodesic. Now, in this gauge the pulsar and Earth (seen as test particles) stay at the same spatial coordinates \cite{Cornish:2009rt}. This implies $(i)$ $\ell_E - \ell_P$ takes the same value with and without GWs and $(ii)$ the proper time measured at Earth is also the coordinate time $t$. Therefore the last term in Eq.~\eqref{eq:t-res1} 
is precisely the sought-after GW-induced timing residual $R_p^{\rm GW}$. Assuming the Earth is at the origin of spatial coordinates, we have 
\beq
R_p^{\rm GW}(t) = \frac12 \hp^a\hp^b \int_{t-D_p}^t dt' ~ h_{ab}(t', (t-t') \hp), \label{eq:R}
\eeq
where $D_p$ is the distance (or time) between Earth and the pulsar. 

It is useful to recast this result in terms of the Fourier transform of the strain. Inserting Eq.~\eqref{eq:Fourier} into the time-residual \eqref{eq:R}, we obtain
\barr
R_p^{\rm GW}(t) &=& \frac12 \hp^a \hp^b \int df \int d^2 \ho~ h_{ab}(f, \ho) \nonumber\\
&& \times  \int_{t-D_p}^t dt' ~ \rme^{2 \pi i f(t' - \hp \cdot \ho (t - t') ) } \nonumber\\
&\equiv& \int df ~ \rme^{2 \pi i f t} R_p^{\rm GW}(f).
\earr
Upon performing the time integral, we find the Fourier transform of the GW-induced time residual $R_p^{\rm GW}(f)$:
\barr
&&R_p^{\rm GW}(f) = \frac{\hp^a \hp^b}{4 \pi i f} \int d^2 \ho~ \frac{h_{ab}(f, \ho)}{(1 + \ho \cdot \hp)}\nonumber\\
&& ~~~~~~~~~~~~~~~~~~~~~~\times \left( 1 - \rme^{-2 \pi i f D_p (1 + \hp \cdot \ho)}\right). \label{eq:R-Fourier}
\earr
The first term in the parenthesis corresponds to the ``Earth term" and the second term to the ``pulsar term".

\subsection{Time-residual correlations}

We define the (one-sided) cross-power spectrum $\mathcal{R}_{pq}^{\rm GW}(f)$ of the GW-induced time residuals at different pulsars $p, q$ as follows:
\beq
\langle R_p^{\rm GW}(f) R_q^{*\rm GW}(f') \rangle = \frac{\delta_{\rm D}(f'-f)}{2} \mathcal{R}_{p q}^{\rm GW}(f). \label{eq:R_pq-def}
\eeq
Using Eq.~\eqref{eq:Pijkl}, we find
\barr
\mathcal{R}_{pq}^{\rm GW}(f) = \frac{1}{(4 \pi f)^2} \int \frac{d^2 \ho}{4 \pi} ~ \frac{\hp^a \hp^b \hq^c \hq^d \bsm{P}_{abcd}(f, \ho) }{(1 + \hp \cdot \ho)(1 + \hq \cdot \ho)} \nonumber\\
\times\left( 1 - \rme^{-2 \pi i f D_p (1 + \hp \cdot \ho )}\right)\left( 1 - \rme^{2 \pi i f D_q (1 + \hq \cdot \ho )}\right). \label{eq:R_pq^GW}
\earr
We can think of the pulsar-term contributions as taking the harmonic transform of the integrand at multipole $\ell \sim 2 \pi f D$ (note that the numerator vanishes as $\ho \rightarrow -\hp$ and $\ho \rightarrow -\hq$ so the integrand is well behaved there). In practice, we have $D \sim $ kpc $\sim 3 \times 10^3$ lightyears and $f \sim $1/yr, thus
\beq
2 \pi f D \approx 2 \times 10^4 \frac{D}{\textrm{kpc}} \frac{f}{\textrm{yr}^{-1}}.
\eeq
Therefore, as long as angular fluctuations of the SGWB on a scale $\ell \gtrsim 10^4$ are negligible, we may safely approximate the terms in parenthesis by 
\beq
\left( 1 - e^{-2 \pi i f D_p (1  + \hp \cdot \ho)}\right) \left( 1 - e^{2 \pi i f D_q (1 + \hq \cdot \ho )}\right) \rightarrow (1+\delta_{pq}), \nonumber
\eeq
where the Kronecker delta accounts for the factor of two if the two pulsars are \emph{identical}, i.e.~have the same location on the sky \emph{and} are at the same distance. See Ref.~\cite{Mingarelli:2018kgp} for an explicit proof of the validity of this approximation for an isotropic SGWB, and \cite{Mingarelli:2014xfa}
for an anisotropic one. 

It will be useful in what follows to introduce some compact notation to denote integrals over the sky. For any two functions of angle on the sky (which from here on we refer to as ``maps'', and represent by bolded symbols throughout) $\bs{\mathcal{M}}_1(\ho)$ and $\bs{\mathcal{M}}_2(\ho)$, we denote for short the scalar product
\beq
\bsm{M}_1 \cdot \bsm{M}_2 \equiv \int \frac{d^2 \ho}{4 \pi} ~\bsm{M}_1(\ho) ~\bsm{M}_2(\ho). \label{eq:dot-prod}
\eeq
Specializing to the total-intensity piece of the SGWB power spectrum, Eq.~\eqref{eq:R_pq^GW} then becomes
\barr
\mathcal{R}_{pq}^{\rm GW}(f) &=& \frac{1 + \delta_{pq}}{(4 \pi  f)^2} \int \frac{d^2 \ho}{4 \pi} \bs{\gamma}_{\hp \hq}(\ho) \bsm{I}(f, \ho)\nonumber\\
&=& \frac{1 + \delta_{pq}}{(4 \pi  f)^2} \bs{\gamma}_{\hp \hq}(\ho) \cdot \bsm{I}(f),
\earr
where we have defined the geometric quantity 
\beq
\bs{\gamma}_{\hp \hq}(\ho) \equiv \frac{\hp^a \hp^b \hq^c \hq^d \mathfrak{I}_{abcd}( \ho) }{(1 + \hp \cdot \ho)(1 + \hq \cdot \ho)} , \label{eq:gamma-def}
\eeq
which can be written explicitly in terms of dot products as follows:
\begin{empheq}[box=\doublebox]{align}
\bs{\gamma}_{\hp \hq}(\ho) = 2\frac{\left( \hat p \cdot \hat q - (\hat p \cdot \ho) (\hat q \cdot \ho)\right)^2}{(1+ \hat p \cdot \ho)(1 + \hat q \cdot \ho)} \nonumber\\
-  (1 - \hat p\cdot \ho)(1 - \hat q\cdot \ho). \label{eq:Gamma}
\end{empheq}
In what follows, we shall refer to $\bs{\gamma}_{\hp \hq}(\ho)$ as the \emph{pairwise timing response function}. It can be expressed in the standard, frame-dependent, notation in terms of the so-called antenna beam patterns $F_{\hp}^+, F_{\hp}^\times$:
\barr
\bs{\gamma}_{\hp \hq}(\ho) &=&  4 \sum_{A = +, \times} F^A_{\hp}(\ho) F^A_{\hq}(\ho),\\
F^A_{\hp}(\ho) &\equiv& \frac12 \frac{\hp^a \hp^b}{1 + \hp \cdot \ho} e^A_{ab}(\ho),
\earr
as can be seen from Eqs.~\eqref{eq:Iabcd-2} and \eqref{eq:gamma-def}. Our geometric approach allowed us to obtain the explicit and clearly frame-independent expression for this function, Eq.~\eqref{eq:Gamma}. The so-called overlap reduction function is then obtained by integrating the angular dependence of the SGWB intensity multiplied by the pairwise timing response function. 

\begin{figure*}
\centering
\includegraphics[width = 2\columnwidth]{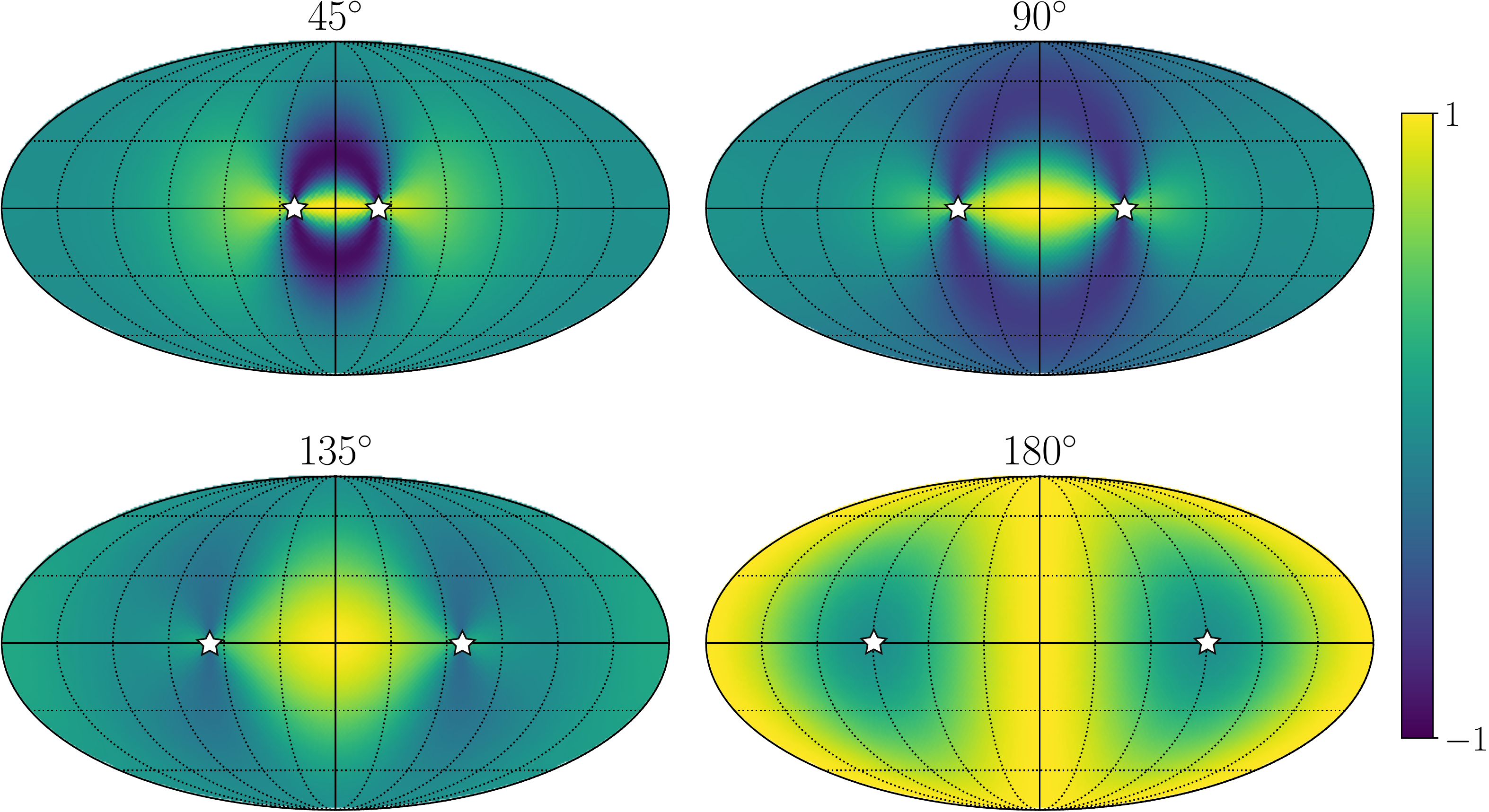}
\caption{Pairwise timing response function $\bs{\gamma}_{\hp \hq}(\ho)$, as function of sky location $\ho$, for four different pulsar separations. The pulsars locations $\hp, \hq$ are shown as stars, and the angle between them is indicated on top of each figure.} \label{fig:gammapq}
\end{figure*}

The kernel $\bs{\gamma}_{\hp \hq}(\ho)$ is symmetric in $\hat p, \hat q$, and so $\mathcal{R}_{pq}(f)$ is symmetric in $p, q$. We show plots of $\bs{\gamma}_{\hp \hq}(\ho)$ in Fig.~\ref{fig:gammapq} as a function of the separation between pulsars $\hp, \hq$ (see also Ref.~\cite{Anholm:2008wy} for similar plots).

As an aside, it is interesting to consider the response function for a single pulsar ($\hp = \hq)$:
\barr
\bs{\gamma}_{\hp \hp}(\ho) &=& (1 - \hp \cdot \ho)^2 \nonumber\\
&=& \frac43  - 2 \hp \cdot \ho +  \hp^a \hp^b \left(\ho_a \ho_b - \frac13 \delta_{ab}\right).
\earr
Therefore a single pulsar is sensitive to a specific linear combination of the SGWB monopole, dipole projected onto $\hp$, and quadrupole twice projected onto $\hp$. Single-pulsar upper limits typically assume an isotropic background. These limits would be weakened if accounting for anisotropies (see, e.g., models 1, 2A-2D considered in Ref.~\cite{Arzoumanian_18}).

\subsection{Response to an isotropic SGWB} \label{sec:mono-dip}

Let us now compute the response to an isotropic SGWB. We define the unit-norm monopole map 
\beq
\bs{1}(\ho) \equiv 1,  \ \ \ \forall~ \ho. 
\eeq
If the SGWB intensity is isotropic, $\bsm{I}(f, \ho) \propto \bs{1}(\ho)$. In that case the time-residual power spectrum $\mathcal{R}_{p q}$ only depends on the scalar product $\mu \equiv \hp \cdot \hq$, with the well-known Hellings and Downs functional dependence \cite{Hellings:1983fr}. We define  
\beq
\mathcal{H}_{\hp \hq} \equiv \bs{1} \cdot \gamma_{\hp \hq} = \int \frac{d^2 \ho}{4 \pi} \bs{\gamma}_{\hp \hq}(\ho) =\mathcal{H}(\hp \cdot \hq).
\eeq
We provide a new and concise derivation of this function in Appendix \ref{app:HD}, making use of our geometric, frame-invariant formalism. With our normalization convention, it is given by
\beq
\mathcal{H}(\mu) = \frac{3 + \mu}{3} + 2 (1 - \mu) \ln\left(\frac{1 - \mu}{2} \right). \label{eq:HD}
\eeq
 
\section{PTA Fisher matrix for the GWB intensity} \label{sec:Fisher}

\subsection{Motivations and general considerations}

The analyses of real PTA data are typically built on a Bayesian framework, and deal directly with the times of arrival (TOAs) of pulsar pulses \cite{Arzoumanian_18, Lentati_15}. The final product of such analyses is to estimate how likely a GWB signal $\bsm{I}(f, \ho)$ is given the data. If the data sample is sufficiently large, the likelihood $\mathcal{L}$ of the GWB intensity $\bsm{I}(f, \ho)$ ought to be approximately Gaussian (see e.g.~Fig.~2 of \cite{Chamberlin_15}), i.e.~\emph{formally} of the form 
\barr
&&-2 \ln \mathcal{L}(\bsm{I}) = \textrm{const} +  \nonumber\\
&&\iint df df'~ [\bsm{I}(f) - \bsm{I}^{\rm ml}(f)] \cdot \bsm{G}_{ff'} \cdot [\bsm{I}(f') - \bsm{I}^{\rm ml}(f')],~~~~ \label{eq:full-likelihood}
\earr
where $\bsm{G}_{ff'}(\ho, \ho')$ is a generalized inverse-covariance ``matrix", and $\bsm{I}^{\rm ml}(f, \ho)$ is the maximum-likelihood SGWB intensity. In full generality, $\bsm{G}_{ff'}$ itself ought to depend on the SGWB intensity (so that the likelihood is not actually Gaussian); nevertheless, we expect that this dependence should only be important once the GWB is \emph{detected} to sufficient significance, as we will quantify shortly. Until then, a \emph{weak-signal} Fisher matrix is sufficiently accurate. 

Our goal here is to provide an \emph{approximate} Fisher matrix, that can be used as a guide to data analysis. This bears similarities with the study of cosmic microwave background (CMB) anisotropies (see e.g.~Ref.~\cite{Tegmark:1996bz}): while the full analysis of CMB data uses a Bayesian framework and deals with the temperature and polarization maps directly, the simple Fisher matrix of CMB \emph{power spectra} allows to make quick and rather accurate detectability forecasts, which serve to inform full data analyses. 

\subsection{Approximate Fisher matrix of band-integrated GWB intensity}

In addition to the stochastic timing residual caused by a SGWB, arrival times are noisy, due to intrinsic pulsar noise and instrumental noise:
\beq
R_p = R_p^{\rm GW} + N_p,  \label{eq:total-Rp}
\eeq
where $N_p$ is the (non-GWB-sourced) timing noise, which we assume to be uncorrelated between pulsars\footnote{A more realistic analysis includes several additional sources of \emph{common} noise, correlated among pulsars, such as global clock errors, or ephemeris errors \cite{Arzoumanian_18}. These additional noise sources do not appear to significantly affect current upper limits on the amplitude of the SGWB \cite{Arzoumanian_18} and we do not include them here. We leave to future work a more detailed treatment including these common noise sources within our Fisher framework.}, and whose power spectrum is $\sigma_p^2(f)$:
\beq
\langle N_p(f) N_q^*(f') \rangle = \delta_{pq} \frac{\delta_{\rm D}(f'-f)}{2} \sigma_p^2(f).\label{eq:sigma_p^2}
\eeq
The standard pulsar analysis fits for several different pulsar-specific sources of noise (e.g., \cite{Chamberlin_15,Hazboun_19}).

In the remainder of this paper, we will work with \emph{band-integrated} quantities: given a frequency bandwidth $\Delta f$, we define the dimensionless band-integrated SGWB intensity
\beq
\bsm{I}_f(\ho) \equiv \int_{f - \Delta f/2}^{f + \Delta f/2} df'~\bsm{I}(f', \ho), \label{eq:I_f}
\eeq
and the band-integrated noise (with dimensions of time squared)
\beq
\sigma_{p, f}^2 \equiv \int_{f - \Delta f/2}^{f + \Delta f/2} df'~ \sigma_p^2(f'). \label{eq:sigma_p-band}
\eeq

We denote by $\mathcal{R}_{pq}(f) = \mathcal{R}_{pq}^{\rm GW}(f) + \delta_{pq} \sigma_p^2(f)$ the total timing-residual cross power spectrum, and by $\mathcal{R}_{pq, f}$ the timing-residual cross-bandpowers (with dimensions of time squared), given by
\barr
\mathcal{R}_{pq, f} &=& \mathcal{R}_{pq, f}^{\rm GW} + \delta_{pq} \sigma_{p, f}^2 \nonumber\\
&=& \frac{1 + \delta_{pq}}{(4 \pi f)^2} ~\bs{\gamma}_{pq}\cdot \bsm{I}_f + \delta_{pq} \sigma_{p, f}^2. \label{eq:I-to-R}
\earr
In what follows, and unless explicitly specified, we always work with band-integrated quantities centered at frequency $f$. To keep the notation manageable, we drop the subscripts $f$ on all band-powers.\\

We label \emph{unique} pairs of \emph{distinct} pulsars by capital indices $I, J, K$. For instance, $I = (p, q) = (q, p)$ represents a unique pair of distinct pulsars $p \neq q$. For $N_{\rm psr}$ pulsars, there are $N_{\rm pair} = N_{\rm psr}(N_{\rm psr}-1)/2$ such distinct pairs. For a pair of distinct pulsars $I$, assuming the SGWB is the only source of correlated noise between distinct pulsars, Eq.~\eqref{eq:I-to-R} simplifies to $\mathcal{R}_{I} =  \bs{\gamma}_I \cdot  \bsm{I}/(4 \pi f)^2$. \\

Let us denote by $\widehat{\mathcal{R}}_{I}$ unbiased estimators of the bandpowers. Let us assume that these estimators are constructed from a large number of effectively uncorrelated samples, implying that they are approximately Gaussian-distributed. Their statistics are thus entirely determined by their $N_{\rm pair} \times N_{\rm pair}$ covariance matrix $\mathcal{C}$, with elements $\mathcal{C}_{IJ}$ (with dimensions of (time)$^4$). Note that this matrix depends on frequency $f$. Under the Gaussian approximation, the joint probability distribution $\mathcal{L}$ of the estimators $\widehat{\mathcal{R}}_{I}$ is therefore 
\barr
&&- 2 \ln \mathcal{L} = \textrm{const}  \nonumber\\
&&+ \sum_{I, J} \left( \widehat{\mathcal{R}}_{I} -  \frac{\bs{\gamma}_{I}  \cdot \bsm{I}}{(4 \pi f)^2}\right)(\mathcal{C}^{-1})_{IJ}\left( \widehat{\mathcal{R}}_{J} -   \frac{\bs{\gamma}_{J}  \cdot \bsm{I}}{(4 \pi f)^2}\right). \label{eq:Fisher-resiudals}
\earr
As is standard in Bayesian data analysis, we view this probability distribution as the likelihood of the signal -- the GWB background bandpower $\bsm{I}(\ho)$ -- given the data. To be precise, this statement assumes a uniform prior on the amplitude of $\bsm{I}(\ho)$. 

In order to write an estimator for the SGWB intensity $\widehat{\bsm{I}}$, we define the \emph{dual maps} $\bs{\gamma}_I^*(\ho)$ (not to be mistaken with complex-conjugates), which are the unique linear combinations of the $\bs{\gamma}_I(\ho)$ satisfying 
\beq
\bs{\gamma}_I^* \cdot \bs{\gamma}_J = \delta_{IJ}.
\eeq
We then define
\beq
\widehat{\bsm{I}}(\ho) \equiv   (4 \pi f)^2 \sum_K \widehat{\mathcal{R}}_{K} \bs{\gamma}_K^*(\ho),
\eeq
which satisfies $(\bs{\gamma}_I \cdot \widehat{\bsm{I}})/(4 \pi f)^2 = \widehat{\mathcal{R}}_I$. We are now finally in the position of defining the Fisher matrix for the bandpowers,
\beq
\boxed{\bsm{F}_f(\ho, \ho') \equiv \frac{1}{(4 \pi f)^4} ~\sum_{I, J}  \bs{\gamma}_I (\ho)  (\mathcal{C}^{-1})_{IJ} \bs{\gamma}_J(\ho')} . \label{eq:Fisher-def}
\eeq
With these definitions, we see that the likelihood for timing residual bandpowers can be rewritten as 
\barr
\mathcal{L} \propto \exp\left\{ - \frac12 \sum_{\textrm{band}(f)} [\bsm{I}_f - \widehat{\bsm{I}}_f] \cdot \bsm{F}_f \cdot [\bsm{I}_f - \widehat{\bsm{I}}_f] \right\}.~~ \label{eq:Fisher-band}
\earr
It might appear at first sight that Eq.~\eqref{eq:Fisher-band} is a probability distribution on the inifinite-dimensional space of maps $\bsm{I}(\ho)$. However, the $N_{\rm pair}$ pairwise-time-residual correlations $\mathcal{R}_{I}$ can only possibly measure $N_{\rm pair}$ projections of the SGWB map. To see what these are precisely, decompose $\bsm{I}(\ho)$ onto a piece which is a linear combination of the functions $\bs{\gamma}_I(\ho)$ -- hence of the $\bs{\gamma}_I^*(\ho)$ -- and a piece which is orthogonal to all of them:
\barr
&&\bsm{I}(\ho) = \bsm{I}_{||}(\ho) + \bsm{I}_{\bot}(\ho),  \nonumber\\
&&\bsm{I}_{||}(\ho) \equiv  (4 \pi f)^2 \sum_K\mathcal{R}_{K} \bs{\gamma}_K^*(\ho),  \\ &&\bsm{I}_{\bot} \cdot \bs{\gamma}_I = 0,  \ \ \ \forall \ I.  \nonumber
\earr
We purposefully denoted by $\mathcal{R}_{K}$ the coefficient of $\bs{\gamma}_K^*$, as it is indeed the time-residual correlation measured by the pulsar pair $K$ (see Eq.~(\ref{eq:I-to-R})). From the expression of $\bsm{F}$, we see that $\bsm{F} \cdot \bsm{I}_{\bot} = 0$. Thus the likelihood function only depends on $\bsm{I}_{||}$, which spans a $N_{\rm pair}$-dimensional space, and contains no information about $\bsm{I}_{\bot}$. Put differently, the components of $\bsm{I}$ orthogonal to the space spanned by the $\bs{\gamma}_I$'s have \emph{infinite noise}. A consequence of this property is that one cannot hope to simultaneously constrain more than $N_{\rm pair}$ statistically independent components of the SGWB map -- be they harmonic coefficients, independent pixels, or any other linear projections. 

\subsection{Weak-signal limit for the Fisher Matrix}

In Appendix \ref{app:radiometer}, we derive the following approximation of the covariance matrix of the estimators for the pairwise-time-residual band powers $\widehat{\mathcal{R}}_{pq}$: for two pairs $I = (p, q)$ and $J = (p', q')$, we have:
\barr
\mathcal{C}_{I J} = \mathcal{C}_{pq, p'q'} &\equiv& \textrm{cov}(\widehat{\mathcal{R}}_{pq}, \widehat{\mathcal{R}}_{p'q'})\nonumber\\
&\approx& \frac1{2 T_{IJ} \Delta f} \left(\mathcal{R}_{pp'} \mathcal{R}_{qq'}+ \mathcal{R}_{pq'} \mathcal{R}_{qp'} \right), \label{eq:C_IJ-general}
\earr
where $T_{IJ}$ is the \emph{effective} total time of observation of the four pulsars $p, q, p', q'$, which we found to be approximately
\beq
T_{IJ} \approx \max\left[ \min(T_p, T_q), \min(T_{p'}, T_{q'}) \right],
\eeq 
if each pulsar $p$ is observed for a total time $T_p$. This equation is a generalization of the radiometer equation for electromagnetic intensity \cite{2017isra.book.....T}, and holds provided the bandwidth $\Delta f$ satisfies
\beq
1/T \ll \Delta f \ll f,
\eeq
where $T$ is the minimum of all observation times. In particular, it only applies for $f \gg 1/T$.

We now specialize to the \emph{weak-signal limit}, i.e.~assume that, for \emph{every pulsar $p$} (and in particular, for the least noisy pulsar), 
\beq
\bsm{I} \ll (4 \pi f)^2\sigma_p^2. \label{eq:weak-signal} 
\eeq
In other words, we assume that the SGWB-induced signal is subdominant to the intrinsic pulsar noise in \emph{each individual pulsar}. 
In this limit, we may approximate $\mathcal{R}_{pq} \approx \delta_{pq} \sigma_p^2$ in the right-hand-side of Eq.~\eqref{eq:C_IJ-general}. As a result, the weak-signal correlation matrix $\mathcal{C}_{IJ}$ is diagonal, and so is its inverse:
\barr
&&\mathcal{C}_{IJ} \approx \frac{\sigma_p^2 \sigma_q^2}{2 T_{pq} \Delta f } \delta_{IJ}, \ \ \ \ \ \ (\mathcal{C}^{-1})_{IJ} \approx \frac{2 T_{pq} \Delta f }{\sigma_p^2 \sigma_q^2} \delta_{IJ}, \label{eq:CIJ-weak}\\
&&I = (p, q), \ \ \ \ \ \ \ T_{pq} \equiv \min(T_p, T_q). \nonumber 
\earr

In addition to stochastic contributions discussed thus far, the timing residual $R_p$ contains a \emph{deterministic} piece, resulting from the pulsar's intrinsic motion, spin down, etc.... To account for these deterministic contributions, a timing model is fitted to pulsars' times of arrival. This process results in a loss of information, quantified by a ``transmission function" $\mathcal{T}_{p}(f)$  \cite{Hazboun_19}. For our purposes, let us note that for all pulsars $\mathcal{T}_{p}(f) \simeq 1$ for $f \gtrsim 1/T$ and $\mathcal{T}_p(f) \simeq (f T_p)^6$ for $f T_p \ll 1$ for most pulsars\footnote{This scaling applies to pulsars with a quadradic spin-down.} \cite{Hazboun_19}. 
In addition (and more relevantly to us for us since we only consider the regime $f T_p \gtrsim 1$), the transmission function filters out harmonics of $1/$year due to degeneracies of timing-model parameters with the motion of the Earth around the Sun. 

Combining Eq.~\eqref{eq:Fisher-def} with Eq.~\eqref{eq:CIJ-weak} and multiplying the contribution of each pair $I = (p, q)$ by $\mathcal{T}_p \mathcal{T}_q$, our final expression for the Fisher matrix for the band-integrated SGWB is therefore 
\begin{empheq}[box=\doublebox]{align}
\bsm{F}_f(\ho, \ho') = \frac1{(4 \pi f)^4} \sum_{p\neq q} \mathcal{T}_p(f) \mathcal{T}_q(f) \nonumber \\
~~~~~\times  \frac{2 T_{pq} \Delta f}{\sigma_{p,f}^2 \sigma_{q,f}^2} ~\bs{\gamma}_{\hp \hq}(\ho) \bs{\gamma}_{\hp\hq}(\ho'). \label{eq:Fisher-final}
\end{empheq}
This weak-signal Fisher matrix is the main result of this paper\footnote{Note that the ``point-spread function" defined in Ref.~\cite{Anholm:2008wy} is proportional to our Fisher matrix, in the case where all pulsars have identical noise.}. It allows us to estimate the signal-to-noise ratio (SNR) of the GWB band-integrated intensity $\bsm{I}_f(\ho)$ with an \emph{arbitrary angular dependence}: \\
\barr
&&\textrm{SNR}^2[\bsm{I}_f] = \bsm{I}_f \cdot \bsm{F}_f \cdot \bsm{I}_f  \nonumber\\ 
&&=  \sum_{p \neq q}~ \mathcal{T}_p(f)\mathcal{T}_q(f) 2  T_{pq} \Delta f ~ \left[ \frac{\bs{\gamma}_{\hp \hq} \cdot \bsm{I}_f}{(4 \pi f)^2\sigma_{p,f} \sigma_{q,f}} \right]^2. \label{eq:SNR2-band}
\earr
Provided the bandwidth is much wider than the inverse of the observation time for each pulsar, $\Delta f \gg  1/T_p$, different bands are uncorrelated, so that the total SNR$^2$ is obtained from summing that of each band:
\barr
&&\textrm{SNR}^2[\textrm{total}] \approx \sum_{\textrm{band}(f)}\textrm{SNR}^2[\bsm{I}_f] \approx 2\sum_{p \neq q} T_{pq} \nonumber\\
&&~~~\times   \int_{1/T_{pq}}^{f_{\max}} df ~\mathcal{T}_p(f)\mathcal{T}_q(f)\left[\frac{\bs{\gamma}_{\hp \hq} \cdot \bsm{I}(f)}{(4 \pi f)^2\sigma_{p}(f) \sigma_{q}(f)} \right]^2, \label{eq:SNR2-total}
\earr
where we replaced the sum over bands by an integral under the assumption that $\Delta f \ll f$. The lower frequency bound is such that $f_{\min} = \max(1/T_p, 1/T_q) = 1/T_{pq}$, and depends on the pulsar pair. The upper frequency bound is the Nyquist frequency $f_{\rm max} = {\rm min}(1/\Delta t_p,1/\Delta t_q)/2$, inversely proportional to the observation cadence. Given the factor $f^{-4}$ in the integrand, unless the SGWB is significantly blue the total SNR is typically dominated by the lowest frequencies, and the upper cutoff has little impact.

Equation \eqref{eq:SNR2-total} generalizes Eq.~(17) of Ref.~\cite{Siemens:2013zla} in several ways. First, it accounts for different observation times for each pulsar. Second, it accounts for the loss of information in the timing-model-fitting process, through the transmission functions $\mathcal{T}_p(f)$. Last but not least, it accounts for an arbitrary angular dependence of the SGWB, rather than assuming a monopole.

Before moving on to applications, let us quantify when the weak-signal limit applies. Suppose all pulsars have a typical observation time $T$ and noise $\sigma_f^2$. Consider moreover frequencies for which $\mathcal{T}_{p}(f) \simeq 1$ (note that for our simple covariance matrix to hold, we require $f \gg 1/T$, thus $\mathcal{T}(f) \simeq 1$ except at harmonics of $1/$year). Eq.~\eqref{eq:SNR2-band} then gives
\beq
\textrm{SNR}^2[\bsm{I}_f(\ho)] \simeq N_{\rm pair} ~ 2 T \Delta f ~ \left(\frac{\mathcal{I}_f}{(4 \pi f)^2 \sigma_{f}^2}\right)^2.
\eeq
The weak-signal approximation \eqref{eq:weak-signal} requires the last term in parentheses to be less than unity. It is thus self-consistent as long as the band-integrated SGWB is detected with a signal-to-noise ratio SNR $\lesssim N_{\rm psr} ~ \sqrt{T \Delta f}$ \emph{in each band}. Unless the SGWB is significantly blue, the total SNR is dominated by the lowest frequencies, so that for the weak-signal limit to be appropriate we must have a total (frequency-integrated) SNR $\lesssim N_{\rm psr}$. 

\section{Idealized case: isotropically distributed identical pulsars} \label{sec:dense-Fisher}

In this section we apply our results to an idealized PTA consisting of $N_{\rm psr} \gg 1$ identical pulsars approximately isotropically distributed on the sky. This limiting case is amenable to analytic approximations, and will serve to cross-check our numerical algorithms when we apply our formalism to real PTAs. We moreover compare our results with those of Ref.~\cite{Hotinli_19}, which apply in this limit.

\subsection{Analytic expression for $N_{\rm psr} \rightarrow \infty$}

\begin{figure}
    \includegraphics[width =\columnwidth]{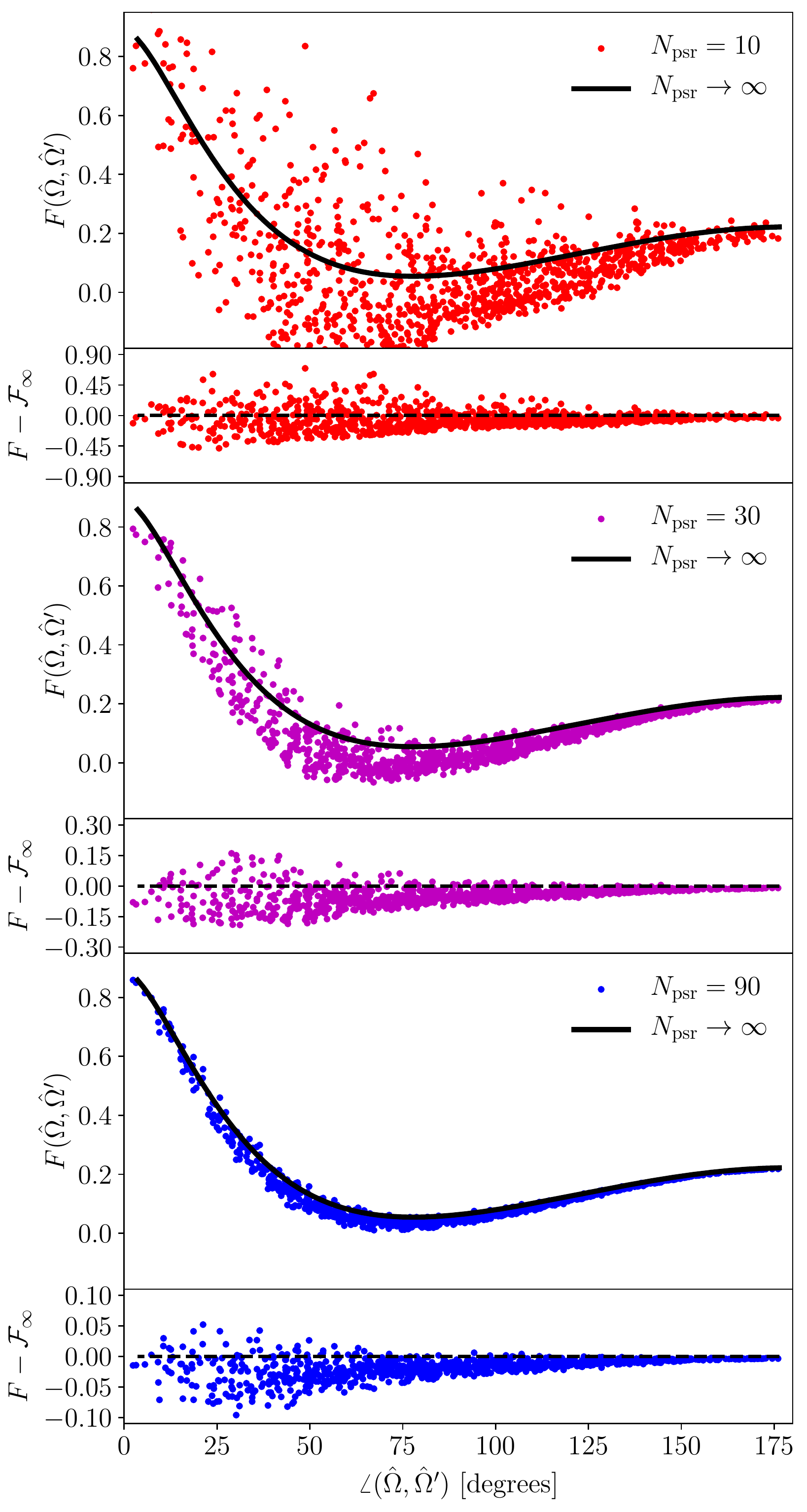}
    \caption{Values of the rescaled Fisher matrix for a finite number of quasi-isotropically distributed identical pulsars (10, 30 and 90, respectively), for 1000 randomly selected pairs of SGWB directions in the sky $(\ho, \ho')$, as a function of the angle between them. The solid black line shows our analytic result, holding for an infinite number of isotropically distributed identical pulsars. We also show the difference between $F$ and its infinite-pulsar limit, $\mathcal{F}_{\infty}$. We see that the difference decreases as $|F - \mathcal{F}_{\infty}| \sim 1/N_{\rm psr}$ (note the different $y$-axis scales in the difference plots).}
    \label{fig:Fdense}
\end{figure}

Suppose all the pulsars have the same noise $\sigma_p = \sigma$, are observed for the same time $T$, and have the same transmission function $\mathcal{T}(f)$. In that limit the Fisher matrix $\bsm{F}$ is given by
\barr
\bsm{F}(\ho, \ho') &=& C ~ \bs{F}(\ho, \ho'), \\
C &\equiv& \frac{\mathcal{T}(f)^2}{(4 \pi f)^4} \frac{2 T \Delta f}{ \sigma^4} N_{\rm pair}, \label{eq:C}\\
\bs{F}(\ho, \ho') &\equiv& \frac1{N_{\rm pair}} \sum_{I} \bs{\gamma}_{I}(\ho)\bs{\gamma}_{I}(\ho'). \label{eq:reduced-F}
\earr
In the limit that $N_{\rm psr} \rightarrow \infty$, assuming the pulsars are isotropically distributed, we find
\barr
\bs{F}(\ho, \ho') &\underset{N_{\rm psr} \rightarrow \infty}{\longrightarrow}& \mathcal{F}_{\infty}(\ho \cdot \ho') \nonumber\\
&\equiv& \int \frac{d^2 \hp}{ 4 \pi} ~ \frac{d^2 \hq}{4 \pi} ~ \bs{\gamma}_{\hp \hq}(\ho)\bs{\gamma}_{\hp \hq}(\ho').
\earr
By symmetry, this is a function of $\chi \equiv \ho \cdot \ho'$ only, which we compute explicitly in Appendix \ref{app:dense-limit}. We derive the following analytic expression:
\barr
\mathcal{F}_{\infty}(\chi) &=& \frac{16}{9 (1 + \chi)^2} \nonumber\\
&&\times \Bigg{[} \left(\frac{1- \chi^2}{4} +  2 - \chi + 3\frac{1 - \chi}{1 + \chi} \log\frac{1-\chi}{2} \right)^2  \nonumber\\
&&~~~~~~ + \left(2-\chi  + 3\frac{1 - \chi}{1 + \chi} \log\frac{1-\chi}{2}\right)^2\Bigg{]}. \label{eq:Fdense}
\earr
We show $\mathcal{F}_{\infty}(\chi)$ as a solid line in Fig.~\ref{fig:Fdense}. For comparison, we also show the reduced Fisher matrix $\bs{F}(\ho, \ho')$ for a finite number of identical, quasi-isotropically distributed pulsars\footnote{To place pulsars quasi-isotropically we arrange them in equal intervals in the azimuthal angle and with the polar angle $\theta = \cos^{-1}(\mathcal{U})$, where $\mathcal{U}$ are a set of uniformly spaced ranging from -1 to 1.}, for 1000 randomly selected pairs of sky directions $(\ho, \ho')$. Only in the limit $N_{\rm psr} \rightarrow \infty$ is the Fisher matrix a function of the angle $\angle(\ho, \ho')$ only; otherwise, it depends on both $\ho$ and $\ho'$, which translates into a scatter of the values of $F(\ho, \ho')$ when plotted as a function of $\angle(\ho, \ho')$. We see that $F(\ho, \ho')$ indeed converges to the function $\mathcal{F}_{\infty}$ as $N_{\rm psr}$ increases, with a difference (both in running mean and scatter) scaling as $\sim 1/N_{\rm psr}$.

The dense-PTA Fisher matrix can be decomposed into Legendre polynomials:
\barr
\mathcal{F}_{\infty}(\ho \cdot \ho') &=& \sum_{\ell} (2 \ell +1)  \mathcal{F}_\ell ~P_{\ell}(\ho \cdot \ho') \nonumber\\
&=& 4 \pi \sum_{\ell,m} \mathcal{F}_{\ell} ~\mathcal{Y}_{\ell m}(\ho) \mathcal{Y}_{\ell m}(\ho'), \label{eq:Legendre}
\earr
where the $\mathcal{Y}_{\ell m}$ are the real spherical harmonics. 

\begin{figure}
    \includegraphics[width = \columnwidth]{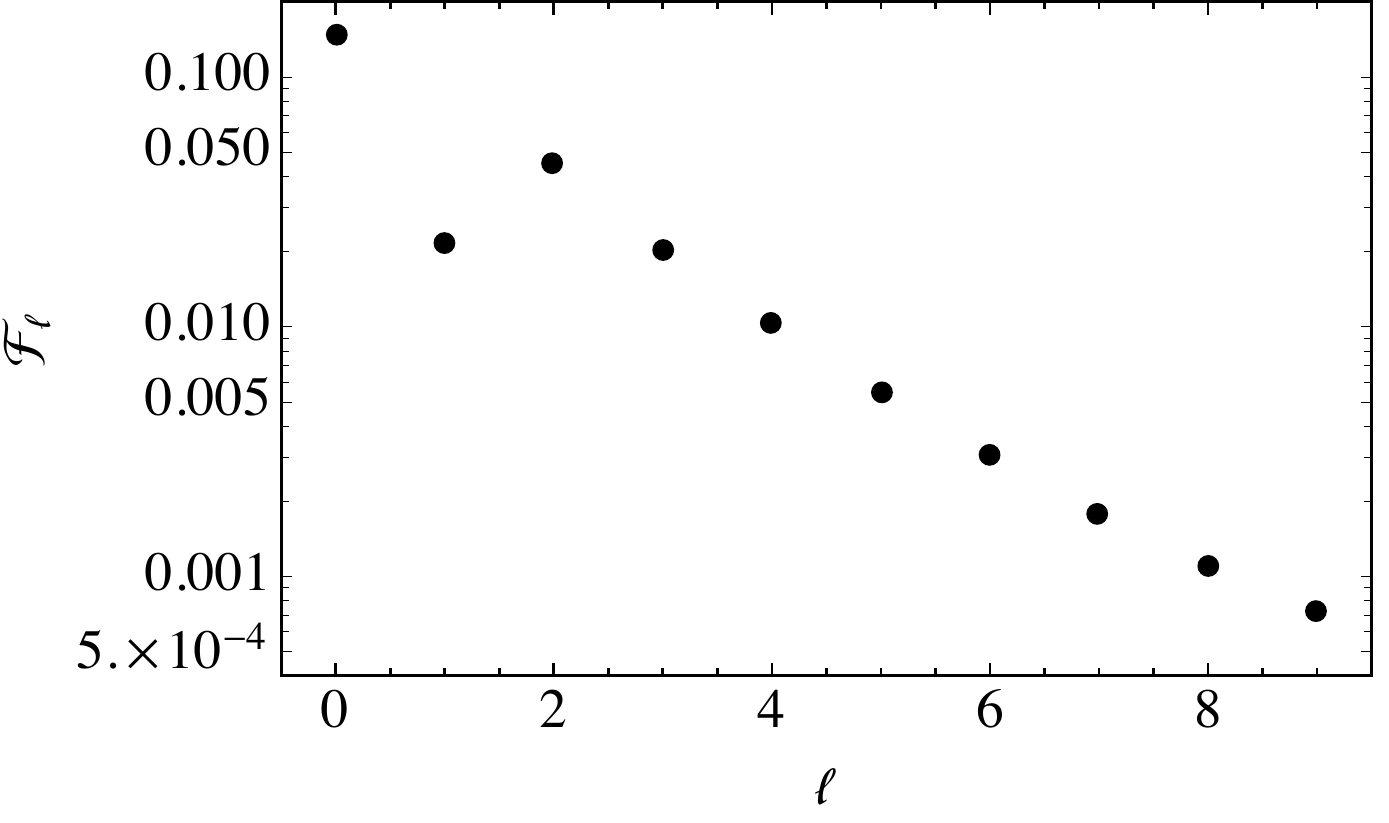}
    \caption{First few Legendre coefficients of the dense-PTA Fisher matrix.} \label{fig:Fdense-leg}
\end{figure}

We show the Legendre coefficients $\mathcal{F}_\ell$ in Fig.~\ref{fig:Fdense-leg}. Interestingly, the amplitude of Legendre coefficients decreases monotonically with $\ell$, \emph{except} for $\mathcal{F}_ 1 \approx \mathcal{F}_0/7$, which is significantly lower than $\mathcal{F}_2$, and comparable to $\mathcal{F}_3$. 

\subsection{Minimum detectable dipolar anisotropy}

Suppose the GWB takes the form 
\beq
\bsm{I} = \mathcal{I}_0 \left(\bs{1} + \sum_{\ell\geq 1, m} g_{\ell m} \bsm{Y}_{\ell m} \right).
\eeq
From Eq.~\eqref{eq:Legendre}, we see that in the limit of a dense array of identical pulsars, the spherical harmonic coefficients of the SGWB are uncorrelated, with noise proportional to $1/\mathcal{F}_\ell$. Explicitly, the signal-to-noise ratio of the coefficients $g_{\ell m}$ is such that
\beq
\frac{\textrm{SNR}^2[\mathcal{I}_0 g_{\ell m} \bsm{Y}_{\ell m}]}{\textrm{SNR}^2[\mathcal{I}_0 \bs{1}]} = \frac{g_{\ell m}^2}{4 \pi} \frac{\mathcal{F}_\ell}{\mathcal{F}_0},
\eeq
where we used the fact that $\bs{1} = \sqrt{4 \pi} \bsm{Y}_{00}$. In particular, in order to detect the $\ell m$-th harmonic with SNR $\geq 3$ requires an amplitude
\beq
g_{\ell m} \geq g_{\ell m, \min} \equiv \sqrt{\frac{\mathcal{F}_0}{\mathcal{F}_{\ell}}} \frac{6 \sqrt{ \pi}}{\textrm{SNR}_0},
\eeq
where SNR$_0 \equiv$ SNR$[\mathcal{I}_0 \bs{1}]$ is the signal-to-noise ratio of the monopole. For the dipole, we find
\beq
g_{1 m, \min} \approx \frac{28}{\textrm{SNR}_0}, \label{eq:g1-min}
\eeq
which is identical to the result of Ref.~\cite{Hotinli_19} in the weak-signal limit.

\subsection{Hot spot in the SGWB}

Now consider a SGWB with a hot spot\footnote{A GW ``beam" in the nomenclature of Ref.~\cite{Hotinli_19}.} in a \emph{known} direction $\ho_0$. Such a hot spot could be generated, for instance, by a concentration of supermassive black hole binaries, sufficiently numerous that the GW background can still be approximated as \emph{stochastic}. Specifically, we assume
\beq
\bsm{I}(\ho) = \mathcal{I}_0 + \mathcal{I}_0 g \left( 4 \pi \delta_{\rm D}(\ho; \ho_0) -1 \right), \label{eq:hot-spot}
\eeq
where we chose the normalization such that the fraction of GW energy density (proportional to the SGWB intensity) in the hot spot is $g$. With this convention, a physical SGWB ought to have $g \leq 1$. 

The joint probability distribution of the monopole and hot spot amplitudes can be obtained from Eq.~\eqref{eq:Fisher-band}, and is a 2-dimensional uncorrelated Gaussian distribution:
\barr
\mathcal{L}(\mathcal{I}_0, \mathcal{I}_0 g) &\propto&  \exp\left[- 2 \pi C \left(\mathcal{I}_0^2 \mathcal{F}_0  + (\mathcal{I}_0 g)^2 (\mathcal{F}_{\infty}(1) - \mathcal{F}_0)\right) \right] \nonumber\\
&=& \exp\left[- \frac{8 \pi}{27} C \left(\mathcal{I}_0^2 + 5 (\mathcal{I}_0 g)^2 \right) \right],
\earr
where the coefficient $C$ is given in Eq.~\eqref{eq:C}, and in the second line we used $\mathcal{F}_0 = 4/27$ and $\mathcal{F}_{\infty}(1) = 8/9$. The variances of the monopole and hot spot amplitudes are thus given by
\beq
\textrm{var}[\mathcal{I}_0] = 5~ \textrm{var}[\mathcal{I}_0 g] = \frac{27}{16\pi} C^{-1}.
\eeq
Hence, for the hot spot to be detectable at the 3-$\sigma$ level, its amplitude needs to be
\barr
g \geq g_{\min} = 3 \frac{\sqrt{\textrm{var}[\mathcal{I}_0 g]}}{\mathcal{I}_0} = \frac{3/\sqrt{5}}{\textrm{SNR}_{0}}, 
\earr
where SNR$_{0} = \mathcal{I}_0/\sqrt{\textrm{var}[\mathcal{I}_0]}$ is again the signal-to-noise ratio of the monopole amplitude. This estimate is in agreement with the numerical result of Ref.~\cite{Hotinli_19} in the weak-signal limit. We thus conclude that, provided with the knowledge of the direction of the hot spot, an idealized PTA would be able to detect a hot spot with amplitude $g \simeq 1$ shortly after the monopole is detected. Without any prior information on the hot spot's direction, of course, this conclusion does not hold.

\subsection{Eigenmaps}

From Eq.~\eqref{eq:Legendre}, we can see that the eigenmaps of the dense-PTA Fisher matrix are the real spherical harmonics. As one can expect, and as we shall see in greater detail in Paper II, the real spherical harmonics are no longer the eigenmaps of realistic PTAs, and therefore do not provide a particularly well adapted basis for searches for anisotropies. To illustrate this, we diagonalize the reduced fisher matrix $\bs{F}(\ho, \ho')$ of an idealized array of a \emph{finite number} of identical, quasi-isotropically distributed pulsars. Specifically, we seek unit-norm maps $\bsm{M}_n(\ho)$ such that 
\beq
\bs{F} \cdot \bsm{M}_n = \frac1{\Sigma_n^2} \bsm{M}_n. \label{eq:eigval}
\eeq
This continuous eigenvalue problem can be transformed into a regular, discrete, eigenvalue problem by seeking $\bsm{M}_n$ as a linear combination of the $\bs{\gamma}_I$:
\beq
\bsm{M}_n(\ho) = \sum_I \mathcal{M}_n^I \bs{\gamma}_I(\ho).
\eeq
The eigenvalue problem \eqref{eq:eigval} is then equivalent to the  discrete $N_{\rm pair} \times N_{\rm pair}$ eigenvalue-problem
\barr
&&\sum_J F_{IJ} \mathcal{M}_n^J = \frac1{\Sigma_n^2} \mathcal{M}_n^I,\\
&&F_{IJ} \equiv \frac{\bs{\gamma}_I \cdot \bs{\gamma}_J}{N_{\rm pair}} . \label{eq:FIJ}
\earr
We thus see that there are exactly $N_{\rm pair}$ principal maps. They do not form a complete set of all possible maps. However, they are a complete set of \emph{observable} maps for a given PTA. Note that the eigenmaps that we derive here are \emph{scalar} maps, corresponding to the intensity of a \emph{stochastic} GW background; this is to be contrasted with the strain eigenmaps derived in Ref.~\cite{Cornish:2014rva}, which apply to \emph{continuous} (i.e.~deterministic) GW searches. There does not appear to be a straightforward connection between our $N_{\rm pair}$ SGWB intensity eigenmaps and the $2 N_{\rm psr}$ strain eigenmaps of Ref.~\cite{Cornish:2014rva}.

\begin{figure}[ht]
    \includegraphics[width = \columnwidth]{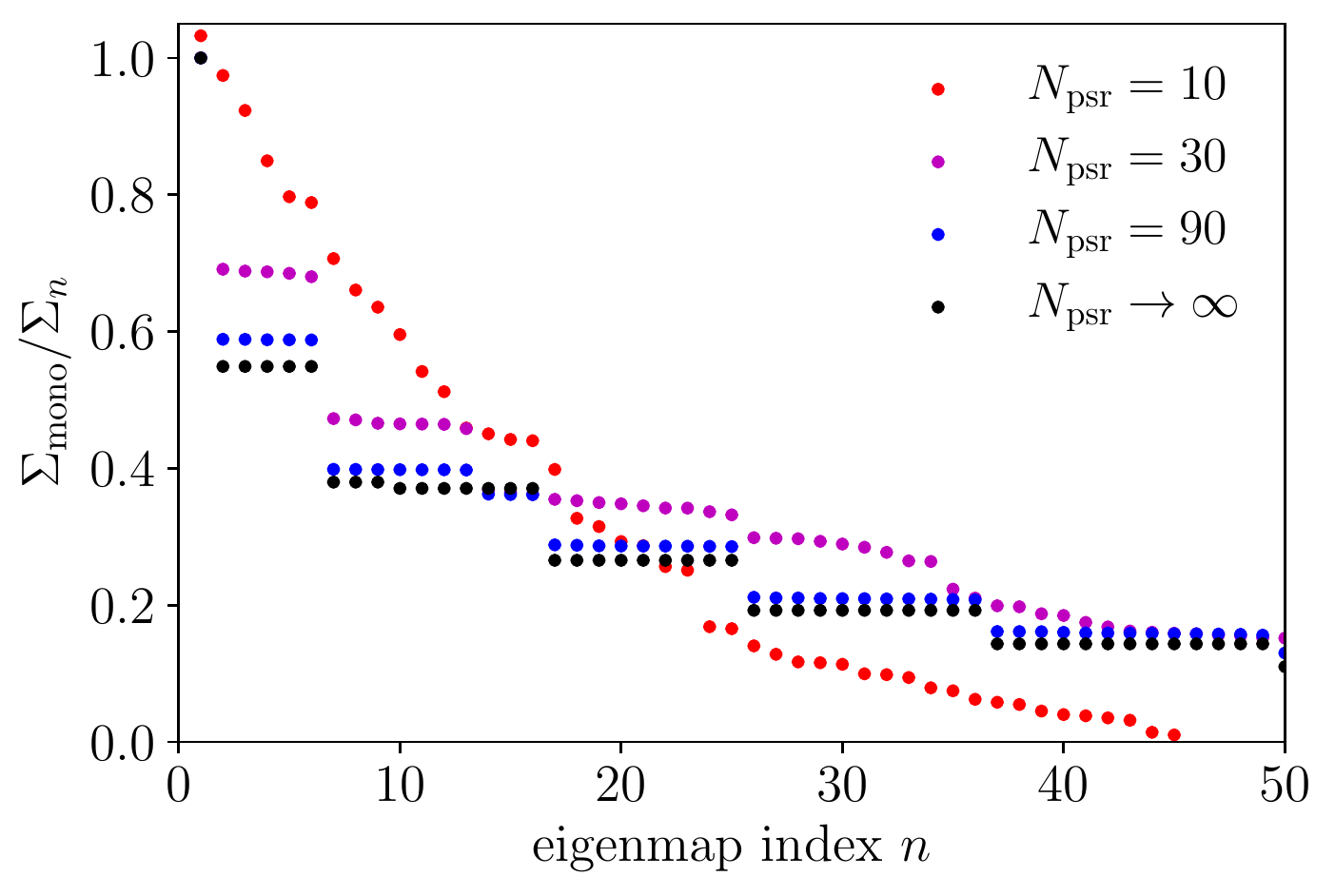}
    \caption{First fifty eigenvalues of quasi-isotropically distributed identical pulsars compared against the dense-pulsar limit $N_{\rm psr} \rightarrow \infty$. The sequences of equal-noise black dots correspond to multipoles $\ell = 0, 2, 1, 3, 4, 5$, in that order. Having a finite number of pulsars perturbs the eigenmaps away from spherical harmonics and breaks the degeneracies in their eigenvalues.}
    \label{fig:iso}
\end{figure}

\begin{figure*}[ht]
\includegraphics[width = 2 \columnwidth]{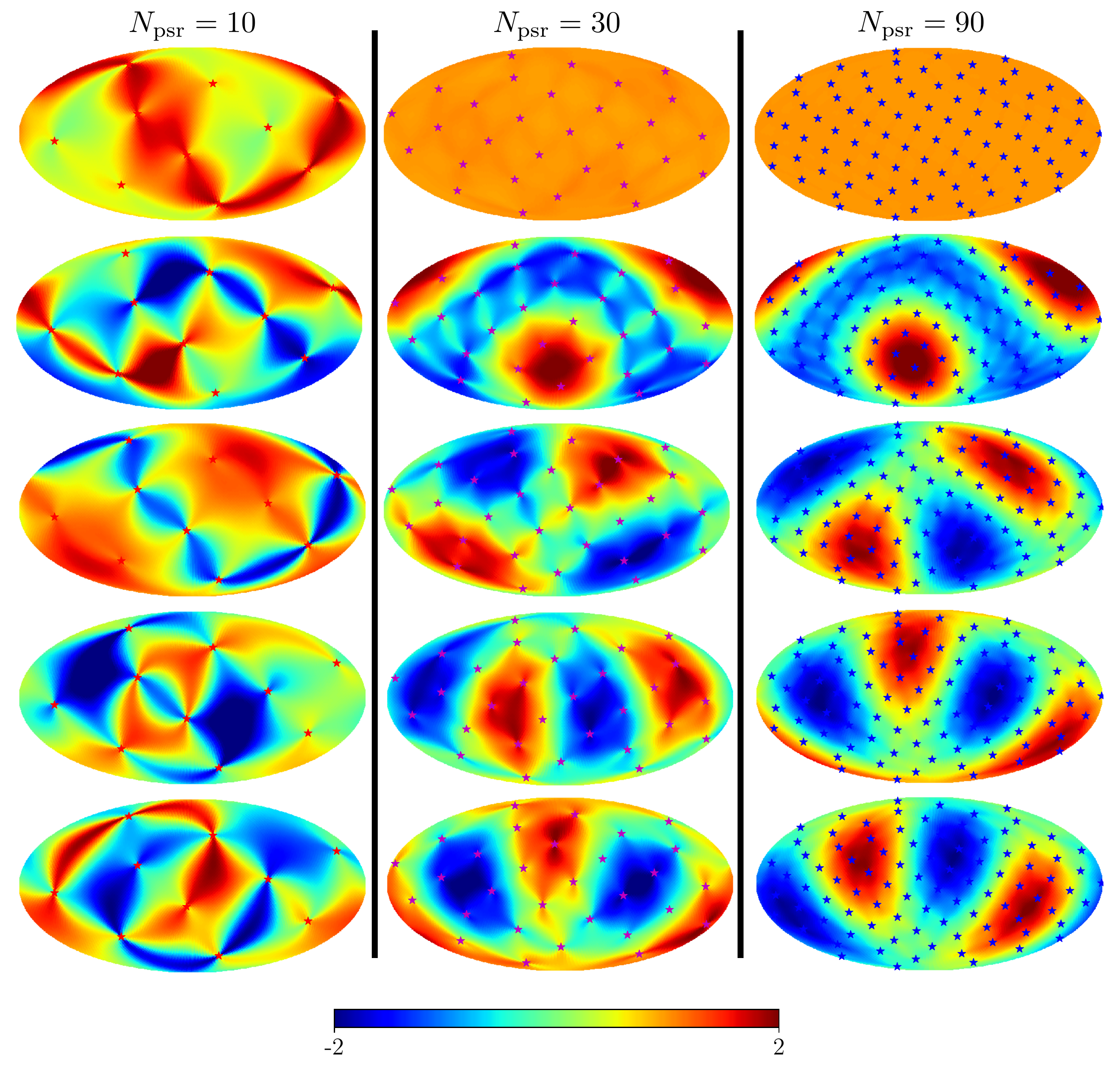}
\caption{First five eigenmaps of the Fisher matrix for $N_{\rm psr} = 10$ (left column), 30 (middle column) and 90 (right column) identical, quasi-isotropically distributed pulsars. As $N_{\rm psr}$ is increased, the first eigenmap approaches the monopole, and the next few eigenmaps become more and more quadrupolar. For $N_{\rm psr} = 10$, the eigenmaps do not resemble spherical harmonics at all. The stars indicate the location of the identical pulsars.} \label{fig:eigmaps-iso}
\end{figure*}

 We show the first 50 eigenvalues Fig.~\ref{fig:iso} for $N_{\rm psr} = 10, 30, 90$, where we compare them against the dense-pulsar limit $N_{\rm psr} \rightarrow \infty$. We see that, as $N_{\rm psr}$ increases, the eigenvalues do converge towards the dense pulsar limit. For $N_{\rm psr} = 90$, one recognizes the sequences of quasi-degenerate eigenvalues, corresponding to the degenerate harmonics for $N_{\rm psr} \rightarrow \infty$. For lower $N_{\rm psr}$, as the Fisher matrix departs further from its $N_{\rm psr} \rightarrow \infty$ limit, eigenmaps ``mix" and are no longer grouped in subsets with similar eigenvalues. This is very similar to the breaking of degeneracy in atomic levels in the presence of a perturbed Hamiltonian. We show the first five eigenmaps in Fig.~\ref{fig:eigmaps-iso}, as a function of $N_{\rm psr}$. We see that as $N_{\rm psr}$ becomes large, the first eigenmap approaches the monopole, and the next two become quadrupolar. For $N_{\rm psr} = 10$, however, the eigenmaps do not at all resemble spherical harmonics. More importantly, as we shall see in Paper II, for realistic pulsar distributions, there exist anisotropies to which a PTA is much more sensitive than the lowest-order spherical harmonics.

\section{Conclusions}

We have derived a band-integrated Fisher matrix for the intensity of a weak, anisotropic SGWB measured by a PTA, Eq.~(\ref{eq:Fisher-final}). This Fisher matrix provides a versatile tool with which we can better study the detectability of anisotropies in the SGWB by PTAs. We derived a simple expression of the SNR of an anisotropic SGWB, Eq.~(\ref{eq:SNR2-total}), generalizing previous results. We moreover derived an exact analytic expression for the Fisher matrix of an idealized PTA consisting of a dense and isotropic distribution of pulsars on the sky. With this matrix, we could recover the results of Ref.~\cite{Hotinli_19} for the detectability of dipolar and hot-spot anisotropies. We illustrated how our formalism is better adapted to realistic PTAs by quantifying the convergence of the Fisher matrix of a finite number of pulsars to that of the dense-pulsar limit. In particular, we showed that, for a finite number of pulsar pairs, the eigenmaps of the Fisher matrix are \emph{not} spherical harmonics, commonly used to study SGWB anisotropies. These $N_{\rm pair}$ eigenmaps best characterize the information content of the Fisher matrix. In a follow-up paper, we will further explore the information content of real PTAs, with unevenly distributed pulsars of unequal noise properties.

In order to arrive at our new Fisher formalism, we re-derived existing results with a fresh look, and presented them in a geometric, coordinate-free form. Let us highlight, in particular, the SGBW power spectrum (a rank-4 tensor) given in Eq.~\eqref{eq:Pijkl2}, and the pairwise timing response function, Eq.~\eqref{eq:Gamma}, which characterizes the correlated response of a pair of pulsars to a generic SGWB intensity map. While in this paper we focused on the total intensity of the SGWB, we have provided all the ingredients needed to extend our results to a circularly or linearly polarized SGWB. Our work could also be generalized to non-Einsteinian polarizations \cite{Eardley:1973br}. Lastly, our Fisher formalism can easily be made more realistic: it can accommodate other sources of correlated pulsar timing residuals, such as global clock errors, and can be generalized to a non-weak SGWB, by using the full expression for the Fisher matrix, Eq.~\eqref{eq:Fisher-def}. Some elements of our Fisher formalism may moreover carry over to other gravitational-wave detection techniques (such as space and ground-based laser interferometers).

The strength of the approach outlined in this paper lies in its ability to clearly and concisely describe the information content of GW measurements. A similar approach for measurements of the CMB \cite{Knox_95} has allowed  accurate, rigorous, and intuitive estimates of the CMB's sensitivity to a variety of effects. At the dawn of GW astronomy, the development of such a tool is both timely and necessary in order to learn as much as we can from the first GW signals that have been and will be measured.  

\section*{Acknowledgements}

We thank Marc Kamionkowski, Selim Hotinli, Jeffrey Hazboun, David Nice, and Joseph Romano for useful discussions. YAH acknowledges support from the National Science Foundation through grant number 1820861. TLS acknowledges support from NASA (though grant number 80NSSC18K0728) and the Research Corporation. The Flatiron Institute is supported by the Simons Foundation.

\onecolumngrid

\begin{appendix}

\section{New derivation of the Hellings and Downs function} \label{app:HD}

We want to compute the following function of $\mu \equiv \hp \cdot \hq$, with $\hp \neq \hq$:
\beq
\mathcal{H}(\mu) \equiv \int \frac{d^2 \ho}{4 \pi} \bs{\gamma}_{\hp \hq}(\ho), 
\eeq
where $\gamma_{\hp \hq}(\ho)$ is given in Eq.~\eqref{eq:Gamma}. Let us define the new variables $
x \equiv \hp \cdot \ho , y \equiv \hq \cdot \ho$, so that 
\beq
\gamma_{\hp \hq} (\ho) = \frac{2(\mu - x y)^2 -(1 - x^2)(1 - y^2)}{(1+x)(1+y)}.
\eeq
The numerator can be rewritten as 
\barr
2(\mu - x y)^2 -(1 - x^2)(1 - y^2) = 2(x^2 + y^2 - 2 \mu xy - 1 + \mu^2) + (1 - x^2) (1 - y^2). \label{eq:numerator}
\earr
The second part simplifies with the denominator and the integral can readily be computed, so we get
\barr
\mathcal{H}(\mu) &=&  \mathcal{J}(\mu) +  (1 + \mu/3), \label{eq:HD-prelim} \\
\mathcal{J}(\mu) &\equiv& 2 \int \frac{d^2 \ho}{4 \pi}  \frac{x^2 + y^2 - 2 \mu xy - (1 - \mu^2) }{(1 + x)(1 + y)}.
\earr
One can show that the coordinates $x, y$ are restricted to the region 
\beq
\mathcal{E}(\mu) = \{ (x, y) ; \ x^2 + y^2 - 2 \mu x y < 1 - \mu^2 \}.
\eeq
The boundary of $\mathcal{E}(\mu)$ is an ellipse whose principal axes are at 45 degree angles from the $(x, y)$ coordinate axes, and with semi-major and semi-minor axes $\sqrt{1\pm \mu}$. Moreover, we can show that the area element is 
\beq
d^2 \ho = \frac{2 d x dy }{\sqrt{1 - \mu^2 - x^2 - y^2 + 2 \mu x y}}. \label{eq:chg-var}
\eeq
With these new variables, the integral $\mathcal{J}(\mu)$ simplifies to
\beq
\mathcal{J}(\mu) = -\int_{\mathcal{E}(\mu)}  \frac{d x dy}{\pi} \frac{\sqrt{1 - \mu^2 - x^2 - y^2 + 2 \mu x y}}{(1 + x)(1 + y)}.
\eeq
For a given $x \in [-1, 1]$, $ y \in [y_-, y_+]$, where the boundaries are given by
\beq
y_{\pm} \equiv \mu x \pm \sqrt{(1 - \mu^2)(1 - x^2)}.
\eeq
We therefore rewrite the integral as
\barr
\mathcal{J}(\mu) = \int_{-1}^1dx \frac{\mathcal{K}(x, \mu)}{1 + x}, 
\earr
where the inner integral is
\barr
\mathcal{K}(x, \mu) \equiv -\frac1{\pi} \int_{y_-}^{y_+} dy \frac{\sqrt{(y_+ - y)(y - y_-)}}{1 + y} = |x + \mu| - (1 + \mu x).
\earr
After performing the simple outer integral, we arrive at
\beq
\mathcal{J}(\mu) = 2 (1 - \mu) \ln\left(\frac{1 - \mu}2\right). 
\eeq
Inserting this result into Eq.~\eqref{eq:HD-prelim}, we finally arrive at the Hellings and Downs function, given in Eq.~\eqref{eq:HD}. 

\section{Probability distribution of timing power spectra} \label{app:radiometer}

In this appendix we derive a simple estimate of the covariance matrix of the pairwise-time-residual cross power spectra. This simple estimate is not meant to follow nor replace a realistic data analysis. Yet, it should provide accurate qualitative scalings, and be quantitatively accurate at the factor-of-few level.

\subsection{Continuous sampling case}
 
Let us suppose that we sample the time residuals $R_p(t)$ of each pulsar $p$ \emph{continuously} over some finite time interval $t \in [-T_p/2, T_p/2]$. Given a frequency $f$, we define
\beq
\tilde{R}_p(f) \equiv \int_{-T_p/2}^{T_p/2} dt ~\rme^{-2\pi i f t} R_p(t)  = T_p \int df_1 R_p(f_1)~ \textrm{sinc}(\pi T_p (f_1 - f)). 
\eeq
The covariance of these quantities is such that
\beq
\langle \tilde{R}_p(f) \tilde{R}_q^*(f') \rangle = \frac12 T_p T_q \int df_1 \mathcal{R}_{pq}(f_1)~ \textrm{sinc}(\pi T_p (f_1 - f))\textrm{sinc}(\pi T_q (f_1 - f')),\label{eq:double-sinc}
\eeq
where $\mathcal{R}_{pq}$ is the total timing residual cross-power spectrum, defined as in Eq.~\eqref{eq:R_pq-def}. Now, assume $\mathcal{R}_{pq}$ varies on a scale $\delta f \sim f$, and that $T_p f, T_q f  \gg 1$. Suppose moreover, for definiteness, that $T_p > T_q$. The sinc function with $T_p$ is narrower, and can be approximated as 
\beq
T_p \textrm{sinc}(\pi T_p (f'-f)) \approx \delta_{\rm D}(f'-f).
\eeq
We define $T_{pq} \equiv \min(T_p, T_q)$. We then get 
\beq
\langle \tilde{R}_p(f) \tilde{R}_q^*(f') \rangle \approx  \frac{T_{pq}}2 \mathcal{R}_{pq}(f) \textrm{sinc}(\pi T_{pq} (f' - f)).
\eeq
Let us now define, for $f > 0$ 
\beq
\widehat{\mathcal{R}}_{pq}(f) \equiv \frac1{T_{pq}}  \left(\tilde{R}_{p}(f) \tilde{R}_{q}^*(f) + \tilde{R}_{q}(f) \tilde{R}_{p}^*(f)\right).
\eeq
From the previous result, $\langle \widehat{\mathcal{R}}_{pq}(f) \rangle = \mathcal{R}_{pq}(f)$, which means that $\widehat{\mathcal{R}}_{pq}$ is an \emph{unbiased estimator} of $\mathcal{R}_{pq}(f)$. Let us now compute its covariance. 
\barr
C_{pq, p'q'}(f, f') &\equiv& \Big{\langle} \left( \widehat{\mathcal{R}}_{pq}(f) - \mathcal{R}_{pq}(f)\right)\left( \widehat{\mathcal{R}}_{p'q'}(f') - \mathcal{R}_{p'q'}(f')\right) \Big{\rangle} = \Big{\langle}\widehat{\mathcal{R}}_{pq}(f) \widehat{\mathcal{R}}_{p'q'}(f') -  \mathcal{R}_{pq}(f)\mathcal{R}_{p'q'}(f')  \Big{\rangle} \nonumber\\
&=& \frac1{2 T_{pq}T_{p'q'}} \Big{\{} T_{pp'} \textrm{sinc}(\pi T_{pp'} (f' - f)) T_{qq'} \textrm{sinc}(\pi T_{qq'} (f' - f))  \mathcal{R}_{pp'}(f) \mathcal{R}_{qq'}(f) \nonumber\\
&& ~~~~~~~~~~~~~ + T_{pq'} \textrm{sinc}(\pi T_{pq'} (f' - f)) T_{qp'} \textrm{sinc}(\pi T_{qp'} (f' - f))  \mathcal{R}_{pq'}(f) \mathcal{R}_{qp'}(f) \Big{\}}.
\earr
We now define 
\beq
T_{\min} \equiv \min(T_{pp'}, T_{qq'}) = \min(T_{pq'}, T_{qp'}) = \min(T_p, T_q, T_{p'}, T_{q'}), \ \ \ \ T_1 \equiv \max(T_{pp'}, T_{qq'}), \ \ \ T_2 \equiv \max(T_{pq'}, T_{qp'}).
\eeq
The broader sinc functions can be evaluated at $f'=f$, and the expression above simplifies to
\beq
C_{pq, p'q'}(f, f') \approx \frac{T_{\min}}{2 T_{pq}T_{p'q'}} \Big{\{}   T_1 \textrm{sinc}(\pi T_1 (f' - f))  \mathcal{R}_{pp'}(f) \mathcal{R}_{qq'}(f) +  T_{2} \textrm{sinc}(\pi T_2 (f' - f))  \mathcal{R}_{pq'}(f) \mathcal{R}_{qp'}(f) \Big{\}}. \label{eq:Cov-ff'}
\eeq
This result shows that the estimators are correlated for frequencies separated by less than $\sim 1/T$, and that their correlation drops for wider frequency separations. 

Let us consider the \emph{bandpowers}, centered at frequencies $f_n = n \Delta f$, where $\Delta f$ is some fixed bandwidth:
\beq
\mathcal{R}_{pq, f_n} \equiv \int_{f_n-\Delta f/2}^{f_n + \Delta f/2} df' \mathcal{R}_{pq}(f').
\eeq
The unbiased estimator $\widehat{\mathcal{R}}_{pq, f_n}$ is obtained by integrating $\widehat{\mathcal{R}}_{pq}(f)$ over a frequency band. Provided $\Delta f/f_n \ll 1$, we have $\mathcal{R}_{pq, f_n} \approx \Delta f \mathcal{R}_{pq}(f_n)$. The covariance of the bandpower estimators is obtained by integrating Eq.~\eqref{eq:Cov-ff'} over the bandwith $\Delta f$ for both frequencies $f, f'$. Provided $T_1 \Delta f, T_2 \Delta f \gg 1$, the sinc functions integrate out, and we are left with
\beq
\textrm{cov}\left(\widehat{\mathcal{R}}_{pq, f_n}, \widehat{\mathcal{R}}_{p'q', f_{n'}} \right) \approx \frac{\delta_{nn'} \Delta f}{2 T_{IJ}} \Big{\{}  \mathcal{R}_{pp'}(f_n) \mathcal{R}_{qq'}(f_n) + \mathcal{R}_{pq'}(f_n) \mathcal{R}_{qp'}(f_n) \Big{\}} \equiv \delta_{nn'} \Delta f ~\mathcal{C}_{I J}(f_n),
\eeq
where the indices $I \equiv (p,q), J \equiv (p',q')$ label pairs of pulsars, and 
\beq
T_{IJ} \equiv \max\left[ \min(T_p, T_q), \min(T_{p'}, T_{q'}) \right].
\eeq

\subsection{Discrete sampling}

Let us now consider the more realistic case where each pulsar $p$ is timed at $(N_p+1) \gg 1$ \emph{discrete} times $t_k = k \Delta t_p$, $k = -N_p/2, ..., N_p/2$, where $T_p = N_p \Delta t_p$. Typically, $\Delta t_p \sim 2-4$ weeks. We now define
\beq
\tilde{R}_p(f) \equiv \Delta t_p \sum_{k = -N_p/2}^{N_p/2} \rme^{- 2 \pi i f t_k} R_p(t_k) = T_p \int df_1 R_p(f_1) \frac{\textrm{sinc}(\pi T_p(f_1 - f))}{\textrm{sinc}( \pi \Delta t_p (f_1 - f))}.
\eeq
The derivation follows the same route as in the continuous case, except for the issue of \emph{aliasing}, translated mathematically by
\beq
T \frac{\textrm{sinc}(\pi T (f'-f))}{\textrm{sinc}( \pi \Delta t (f'-f))} \approx \sum_{n = -\infty}^{\infty} (-1)^n \delta_{\rm D}(f'- f - n/\Delta t).
\eeq
If the timing cross-power spectrum $\mathcal{R}_{pq}(f)$ scales as $f^{-\alpha}$, with $\alpha > 1$, then aliasing does not affect any of the results above, as the contribution from higher-order multiples of $1/\Delta t_p$ is negligible relative to the fundamental mode $n = 0$. This is expected to be the case for $p \neq q$. However, the single-pulsar timing residual power spectrum $\mathcal{R}_{pp}(f)$ has a constant white noise piece $P_p(f) = \sigma_{p,\rm wn}^2 t_{\rm obs} $ at sufficiently high frequencies, up to a maximum frequency $|f_{\max}| = 1/t_{\rm obs}$. Here $t_{\rm obs}$ is the duration of an \emph{individual observation} (typically, $t_{\rm obs} \sim 30$ minutes), from which a single, averaged ``time of arrival" (TOA) is obtained, and $\sigma_{p, \rm wn}^2$ is the variance of the timing residual (after fitting a timing model) between individual observations. Thus, we find 
\beq
\langle \tilde{R}_p(f) \tilde{R}_p^*(f') \rangle \approx \frac{T_p}{2} \sum_{n = -\Delta t_p/t_{\rm obs}}^{\Delta t_p/t_{\rm obs}} \mathcal{R}_{pp}(f - n/\Delta t_p) \textrm{sinc}(\pi T_p(f'-f)) = \frac{T_p}{2} \textrm{sinc}(\pi T_p(f'-f)) \left(\mathcal{R}_{pp}(f) + 2 \sigma_{p, \rm wn}^2 \Delta t_p \right).
\eeq
Hence, the results obtained for the continuum-sampling case carry over to the discrete-sampling case, provided one includes the white noise contribution $2 \sigma_{p, \rm wn}^2 \Delta t_p$ in pulsar autocorrelations. We emphasize that this term accounts for aliasing, i.e.~from the white noise power at all harmonics of $1/\Delta t_p$, up to the maximum frequency $1/t_{\rm obs}$.

\section{Dense and isotropic pulsar distribution limit} \label{app:dense-limit}

In the limit where pulsars are densely and isotropically distributed across the sky, the Fisher matrix becomes proportional to 
\beq
\mathcal{F}_{\infty}(\chi) \equiv \int \frac{d^2 \hp}{4 \pi} \frac{d^2 \hq}{4 \pi} \bs{\gamma}_{\hp \hq}(\ho) \bs{\gamma}_{\hp \hq}(\ho'),  \ \ \ \ \ \ \chi \equiv \ho \cdot \ho'. \label{eq:Fdense-app}
\eeq
Now remember that the pairwise timing response function is given by 
\beq
\gamma_{\hp \hq}(\ho) = \frac{\hp^a \hp^b \hq^c \hq^d \mathfrak{I}_{abcd}(\ho)}{(1 + \hp \cdot \ho)(1 + \hq \cdot \ho)}.
\eeq
The double angular integral over pulsar directions can thus be factorized:
\barr
\mathcal{F}_{\infty}(\ho \cdot \ho') &=& \bsm{\tilde{K}}_{aba'b'}(\ho, \ho') \mathfrak{I}_{a'b'c'd'}(\ho') \bsm{\tilde{K}}_{c'd' cd}(\ho', \ho)  \mathfrak{I}_{cd ab}(\ho) ,\\
\bsm{\tilde{K}}_{aba'b'}(\ho, \ho') &\equiv& \int  \frac{d^2 \hp}{4 \pi} \frac{\hp_a \hp_b \hp_{a'} \hp_{b'}}{(1 + \hp \cdot \ho)(1 + \hp \cdot \ho')}.
\earr
Since $\mathfrak{I}_{a'b'c'd'}(\ho')$ is orthogonal to $\ho'$ in all indices, and trace-free in the first and last pairs of indices, one may replace $\bsm{\tilde{K}}_{aba'b'}(\ho, \ho')$ by its projection orthogonal to $\ho'$ and trace free on the right two indices. The same holds for the left two indices. Upon projecting on $\bsm{I}$, we find
\barr
\mathcal{F}_{\infty}(\ho \cdot \ho') &=& 4 ~\bsm{K}_{aba'b'}(\ho, \ho') \bsm{K}_{aba'b'}(\ho, \ho'), \\
\bsm{K}_{aba'b'}(\ho, \ho') &\equiv& \int  \frac{d^2 \hp}{4 \pi} \frac{(\hp^{\bot}_a \hp^{\bot}_b - \frac12 (\hp^{\bot})^2 \delta_{ab}^{\bot}) (\hp_{a'}^{\bot'} \hp_{b'}^{\bot'} - \frac12 (\hp^{\bot'})^2 \delta_{a' b'}^{\bot'})}{(1 + \hp \cdot \ho)(1 + \hp \cdot \ho')},
\earr
where $\hp^{\bot} \equiv \hp - (\hp \cdot \ho) \ho $ and $\hp^{\bot'} \equiv \hp - (\hp \cdot \ho') \ho'$.

The tensor $\bsm{K}_{aba'b'}(\ho, \ho')$ is symmetric, trace-free and orthogonal to $\ho$ in its first two indices, and symmetric, trace-free and orthogonal to $\ho'$ in its last two indices. It therefore has 4 independent components. 

Given the preferred directions $\ho, \ho'$, one may construct two rank-2 tensors that are symmetric, trace-free and orthogonal to $\ho$ on both indices. Defining $\bs{V} = \ho \times \ho'$, those two tensors are
\beq
\bsm{A}(\ho, \ho') \equiv (\ho' - \chi \ho) \otimes (\ho' - \chi \ho)  - \bs{V} \otimes \bs{V} , \ \ \ \ \ \bsm{B}(\ho, \ho') \equiv (\ho' - \chi \ho) \otimes \bs{V} + \bs{V} \otimes  (\ho' - \chi \ho).
\eeq
Note that both $\ho' - \chi \ho$ and $\bs{V}$ have norm $\sqrt{1 - \chi^2}$, where $\chi \equiv \ho \cdot \ho'$, which is why $\bsm{A}$ is indeed trace-free.

Therefore the rank-4 tensor $\bsm{K}(\ho, \ho')$ ought to take the form:
\beq
\bsm{K}(\ho, \ho') = A \bsm{A}(\ho, \ho') \otimes \bsm{A}(\ho', \ho) + B  \bsm{B}(\ho, \ho') \otimes \bsm{B}(\ho', \ho) + C \bsm{A}(\ho, \ho') \otimes \bsm{B}(\ho', \ho) + D \bsm{B}(\ho, \ho') \otimes \bsm{A}(\ho', \ho),  
\eeq
where $A, B, C, D$ only depend on $\chi$. Now, $\bsm{K}$ is symmetric under exchange of the first two indices and last two indices, simultaneously with exchange of $\ho, \ho'$. Since $\bsm{B}(\ho', \ho) = - \bsm{B}(\ho, \ho')$ (if we do not change the definition of $\bs{V} = \ho \times \ho'$), then we must have $D = - C$. Lastly, $\bsm{K}(- \ho, -\ho') = \bsm{K}(\ho, \ho')$, which implies $C = D = 0$. We have thus found that 
\beq
\bsm{K}(\ho, \ho') = A(\chi) ~ \bsm{A}(\ho, \ho') \otimes \bsm{A}(\ho', \ho) + B(\chi)~  \bsm{B}(\ho, \ho') \otimes \bsm{B}(\ho', \ho)
\eeq
The desired function is the contraction of $\bsm{K}$ with itself in its first two indices and in its last two indices. Using the fact that $(\bsm{A}:\bsm{B}) \equiv \mathcal{A}_{ab} \mathcal{B}_{ab} = 0$, we get
\beq
\mathcal{F}_{\infty}(\chi) = 4 \left[ A^2 (\bsm{A}:\bsm{A})^2 + B^2 (\bsm{B}:\bsm{B})^2 \right].
\eeq
Lastly, we have 
\beq
\bsm{A}:\bsm{A} = 2(1 - \chi^2)^2 = \bsm{B}:\bsm{B}.
\eeq
Hence we have found
\beq
\mathcal{F}_{\infty}(\chi) = 16 (1 - \chi^2)^4 \left[A^2 + B^2 \right].
\eeq
The next step is now to determine $A(\chi)$ and $B(\chi)$. We do so by computing the following contractions:
\barr
(\ho' \otimes \ho') : \bsm{A}(\ho, \ho') = (1 - \chi^2)^2 = (\ho' \otimes \bs{V}) : \bsm{B}(\ho, \ho'), \\
(\ho' \otimes \bs{V}) : \bsm{A}(\ho, \ho') = 0 = (\ho' \otimes \ho') : \bsm{B}(\ho, \ho')
\earr
We therefore arrive at
\barr
(1 - \chi^2)^4 A = (\ho' \otimes \ho') : \bsm{K}(\ho, \ho') : (\ho \otimes \ho), \\
(1 - \chi^2)^4 B = (\ho' \otimes \bs{V}) : \bsm{K}(\ho, \ho') : (\bs{V} \otimes \ho).
\earr
It is now ``only a matter of" computing these contractions, which are scalar integrals. To do so, let us introduce some notation:
\beq
x \equiv \hp \cdot \ho, \ \ \ \ y \equiv \hp \cdot \ho', \ \ \ \ P(x, y, \chi) \equiv 1 - \chi^2 - x^2 - y^2 + 2 \chi x y = (\bs{V} \cdot \hat{p})^2 \geq 0.
\eeq
We then get
\barr
\ho'_a \ho'_b \left(\hp^{\bot}_a \hp^{\bot}_b - \frac12 (\hp^{\bot})^2 \delta_{ab}^{\bot}\right) &=& (y - \chi x)^2 - \frac12 (1 - \chi^2)(1 - x^2)  =  \frac12 (1 - \chi^2) (1 - x^2) - P(x, y, \chi),\\
\ho'_a V_b \left(\hp^{\bot}_a \hp^{\bot}_b - \frac12 (\hp^{\bot})^2 \delta_{ab}^{\bot}\right) &=& (y - \chi x) \bs{V} \cdot \hp = \pm (y - \chi x) \sqrt{P(x, y, \chi)}.
\earr
So we find  
\barr
(1 - \chi^2)^4 A(\chi) &=& \int \frac{d^2 \hp}{4 \pi} \frac{\left[\frac12 (1 - \chi^2) (1 - x^2) - P(x, y, \chi)  \right]\left[\frac12 (1 - \chi^2) (1 - y^2) - P(x, y, \chi) \right]}{(1 + x) (1 + y)} \nonumber\\
(1 - \chi^2)^4 B(\chi) &=& \int \frac{d^2 \hp}{4 \pi} \frac{(y - \chi x)(x - \chi y) P(x, y, \chi)}{(1 + x)(1 + y)}.
\earr
Now recall, from Appendix \ref{app:HD}, that 
\beq
d^2 \hp = \frac{2 d x dy}{\sqrt{P(x, y, \chi)}}.
\eeq
Evaluating the integrals, and simplifying, we find
\barr
(1 - \chi^2)^4 A(\chi) &=& \frac{(1 - \chi)^2}{3} \left( \frac{1 +\chi}{4}( 9 - \chi ( 4 + \chi)) + 3(1 - \chi) \log\frac{1-\chi}{2} \right), \\
(1 - \chi^2)^4 B(\chi)  &=& \frac{(1 - \chi)^2}{3} \left((\chi +1)(\chi - 2) - 3(1 - \chi) \log\frac{1-\chi}{2}\right).
\earr 
After simplifying, we thus arrive at our final expression, Eq.~\eqref{eq:Fdense}.

\end{appendix}

\twocolumngrid

\bibliography{PTA-Fisher.bib}

\end{document}